\documentclass[fleqn,usenatbib,useAMS]{mnras}

\usepackage{graphicx}	
\usepackage{amsmath}	
\usepackage{amssymb}	
\usepackage{multicol}        
\usepackage{bm}		
\usepackage{pdflscape}	
\usepackage{xcolor}
\usepackage{ulem}
\usepackage{caption}
\usepackage{subcaption}
\usepackage{longtable}
\usepackage{supertabular}

\usepackage[T1]{fontenc}
\usepackage{ae,aecompl}
\usepackage{newtxtext,newtxmath}

\defcitealias{Jauzac2016}{J16}
\defcitealias{Niemiec2023}{N23}

\newcommand{\hst}{\textit{HST}}
\newcommand{\jwst}{\textit{JWST}}
\newcommand{\lenstool}{\textsc{Lenstool}}
\newcommand{\pyrrg}{\textsc{pyRRG}}

\title[SHMF for Abell\,2744 with strong+weak lensing]{The BUFFALO Survey : The Subhalo Mass Function for Abell\,2744 with Strong+Weak Gravitational Lensing}

\author[N. R. Patel et al. 2026]
{Nency R. Patel,$^{1,2}$\thanks{Email : nency.r.patel@durham.ac.uk},
Mathilde Jauzac,$^{3,1,2,4,5}$ 
Anna Niemiec,$^{6}$
Guillaume Mahler,$^{7}$ 
David Lagattuta,$^{8}$
\newauthor
David Harvey,$^{9}$
Johannes Schwinn,$^{10,2}$
Andrew Robertson,$^{11}$
Lukas J. Furtak,$^{12,13}$
Sownak Bose,$^{2}$
\newauthor
Alastair Edge$^{1,2}$,
Anton Koekemoer$^{14}$,
Jose Diego$^{15}$,
Andreas Faisst$^{16}$,
Eric Jullo$^{17}$,
Marceau Limousin$^{17}$,
\newauthor
Richard Massey$^{1,2}$,
Derek Perera$^{18}$,
and Sut-Ieng Tam$^{19}$
\\
\\
$^{1}$Centre for Extragalactic Astronomy, Department of Physics, Durham University, Durham DH1 3LE, UK \\
$^{2}$Institute for Computational Cosmology, Department of Physics, Durham University, Durham DH1 3LE, UK\\
$^{3}$Univ Toulouse, CNES, CNRS, IRAP, Toulouse, France\\
$^{4}$Astrophysics Research Centre, University of KwaZulu-Natal, Westville Campus, Durban 4041, South Africa \\
$^{5}$School of Mathematics, Statistics \& Computer Science, University of KwaZulu-Natal, Westville Campus, Durban 4041, South Africa \\
$^{6}$ Univ. Grenoble Alpes, CNRS, Grenoble INP, LPSC-IN2P3, 38000 Grenoble, France\\
$^{7}$ STAR Institute, Quartier Agora - All\'ee du six Ao\^ut, 19c B-4000 Li\`ege, Belgium \\
$^{8}$ Centre for Astrophysics Research, Department of Physics, Astronomy and Mathematics, University of Hertfordshire, Hatfield AL10 9AB, UK \\
$^{9}$ Laboratoire d’Astrophysique, EPFL, Observatoire de Sauverny, 1290 Versoix, Switzerland \\
$^{10}$ Universit\"at Heidelberg, Zentrum f\"ur Astronomie, Institut f\"ur Theoretische Astrophysik, Philosophenweg 12, 69120 Heidelberg, Germany\\
$^{11}$ Carnegie Observatories, 813 Santa Barbara Street, Pasadena, CA 91101, USA \\
$^{12}$ Cosmic Frontier Center, The University of Texas at Austin, Austin, TX 78712, USA \\
$^{13}$ Department of Astronomy, The University of Texas at Austin, Austin, TX 78712, USA \\
$^{14}$ Space Telescope Science Institute, 3700 San Martin Drive, Baltimore, MD 21218, USA \\
$^{15}$ Instituto de F´ısica de Cantabria (CSIC-UC), Avda. Los Castros s/n, 39005 Santander, Spain\\
$^{16}$ Caltech/IPAC, 1200 E. California Boulevard, Pasadena, CA 91125, USA \\
$^{17}$ Aix Marseille Univ, CNRS, CNES, LAM, Marseille, France \\
$^{18}$ School of Physics and Astronomy, University of Minnesota, Minneapolis, MN, 55455, USA \\
$^{19}$ Institute of Physics, National Yang Ming Chiao Tung University, No. 1001, Daxue Rd. East Dist., Hsinchu City, Taiwan\\
}

\date{Accepted XXX. Received YYY; in original form ZZZ}

\pubyear{2026}

\begin{document}
\label{firstpage}
\pagerange{\pageref{firstpage}--\pageref{lastpage}}
\maketitle

\begin{abstract}
We present a combined strong and weak gravitational lensing mass model of the massive merging galaxy cluster Abell\,2744 ($z = 0.308$). The mass reconstruction combines strong-lensing constraints in the cluster core with weak-lensing measurements derived from deep imaging with the \textit{Hubble Space Telescope (HST)} and the \textit{James Webb Space Telescope (JWST)}, enabling a reconstruction of the projected mass distribution from the core to the cluster outskirts. Using the resulting mass map, we identify cluster substructures and construct the subhalo mass function (SHMF) for Abell\,2744. To interpret the observed substructure population, we apply identical detection methods to projected mass maps of massive galaxy clusters from the BAryons and HAloes of MAssive Systems (BAHAMAS) simulations. We observe a systematic excess of detected substructures in Abell\,2744 compared to BAHAMAS. We interpret this excess as likely reflecting Abell\,2744's exceptionally disturbed dynamical state, characterised by complex ongoing mergers, rather than tension with $\Lambda$Cold dark matter ($\Lambda$CDM) or Self-interacting dark matter (SIDM) predictions. Although, confirming this interpretation will require extending the analysis to a larger sample of clusters spanning a range of dynamical states.

\end{abstract}





\section{INTRODUCTION}
Galaxy clusters represent the most massive gravitationally bound structures in the Universe and form at the nodes of the cosmic web through 
a hierarchical scenario \citep{Kravtsov2012}. Their assembly is driven primarily by the accretion and merging of smaller structures, making them powerful laboratories to test the nature of dark matter. In particular, massive merging clusters probe a regime where dark matter dominates the gravitational potential while baryonic effects are comparatively subdominant, allowing deviations from the predictions of collisionless cold dark matter (CDM) to be more readily identified \citep{Allen2011}.

During major merger events, the cluster potential is driven far from equilibrium, producing pronounced substructures in the galaxy and dark matter distributions. These substructures, often referred to as subhalos, are smaller gravitationally bound dark matter structures embedded within the larger cluster halo and are typically associated with infalling galaxy groups or cluster member galaxies. They retain memory of the underlying dark matter physics through their abundance, internal mass profiles, and spatial distribution. As a result, merging clusters provide a uniquely sensitive environment for investigating dark matter properties such as self interaction cross sections, velocity dependence, and departures from collisionless behaviours \citep{Markevitch2004,Clowe2006,Harvey2015}. Observational signatures of such effects have been explored through offsets between dark matter, stellar, and gaseous components, as well as through the survival and disruption of substructures during mergers \citep{Randall2008}.

A particularly powerful statistical probe of dark matter in this context is the subhalo mass function (SHMF), which quantifies the abundance of gravitationally bound substructures as a function of their mass within a host halo. Numerical simulations predict that the SHMF is largely universal in CDM cosmologies, following a quasi-power law form over a wide mass range \citep{Springel2008,Gao2012}. However, alternative dark matter scenarios, including self interacting dark matter (SIDM) and warm dark matter (WDM) models, can suppress or modify the SHMF by altering subhalo survival, tidal stripping efficiency, and central density profiles \citep{Vogelsberger2012,Rocha2013,Bhattacharyya2022,Sirks2022}. Measuring the SHMF in massive clusters therefore provides a direct test of dark matter physics.


Gravitational lensing represents a unique method to probe the projected mass distribution of galaxy clusters and the substructures they are composed of. The deflection of light from a background source caused by an intervening mass is called gravitational lensing. It can be observed mainly in two regimes, each providing different observational signatures: the strong and the weak lensing regimes. 
Strong gravitational lensing is seen in the high density regions such as the core of a cluster \citep{Narayan1984,Lynds1987,Soucail1987,Soucail1988,Fort1994,NatarajanKneib1997,kneib2011,Natarajan2006,Natarajan2024}. In this regime, lensing strongly constrains the inner mass distribution of clusters on scales of tens to hundreds of kiloparsecs. The high density deflects the light strongly, which results in large distortions, giant arcs and multiple images of the background source, observable in high resolution images.
Weak lensing on the other hand occurs in regions of low density such as the outskirts of clusters \citep{Tyson1990,Kaiser1993,Squires1996,bartelmann2001,Hoekstra2013,Mandelbaum2018}. Weak lensing constrains the large scale and diffuse mass distribution of clusters, extending from intermediate radii out to the virial radius and beyond. The light from the background source will be weakly deflected forming minor distortions, called shear, in the images of the lensed sources. It is nearly impossible to differentiate between the intrinsic ellipticity and the shear induced by weak lensing for an individual source. As a result, weak lensing is quantified statistically.
When combined, strong and weak lensing enable the detection and characterisation of subhalos over a broad range of masses and scales \citep[up to the virial radius of the cluster,][]{bradac2005,merten2009,Jauzac2016,Umetsu2018} making lensing-based SHMF measurements particularly well suited for studies of merging clusters \citep{Natarajan2017,Meneghetti2020}.

This paper focuses on one of the most fascinating and complex galaxy clusters in the observable Universe, Abell\,2744 (also known as AC\,118, MACS\,J0014.3-3022 and the ``Pandora's Cluster''). At a redshift of $z=0.308$, and with a total projected mass of approximately $2.6 \times 10^{15} M_\odot$, Abell\,2744 ranks among the most massive galaxy clusters discovered to date. Due to its disturbed dynamical state and powerful lensing capabilities, the cluster has been extensively studied \citep{Jauzac2015_a2744,Jauzac2016,Schwinn2017,schwinn2018,Jauzac2018,Mahler2018,Furtak2023b,Harvey2024,Cha2024}.

Abell\,2744 is an X-ray luminous cluster \citep[$L_{\mathrm{X}} = 3.1 \times 10^{45}~\mathrm{erg}~\mathrm{s}^{-1}$][]{Allen1998}, and was first identified as a major merger through radio observations of its extended halo~\citep{Giovannini1999,Govoni2001}. X-ray studies confirmed this merger scenario and revealed complex substructures~\citep{Kempner2004,Zhang2004,Owers2011}, while kinematic analyses revealed a high global line-of-sight velocity dispersion of $\sigma_V = 1770\,km\,s^{-1}$ and measured a bimodal velocity distribution in the cluster core, together with a third group of members coincident with the north-western X-ray peak, for a total of three galaxy 
groups~\citep{Girardi2001,Boschin2006,Braglia2007}.

Gravitational lensing studies began with~\cite{Smail1997} who identified strong-lensing features using \hst~data. \cite{Allen1998} found discrepancies between the X-ray and lensing masses, supporting the merger interpretation. \cite{merten2011} conducted the first comprehensive multi-wavelength analysis of the cluster, combining strong lensing, weak lensing, and X-ray data. \citet{merten2011} confirms the presence the North and the North-West substructures in the cluster. 
After the \textit{Hubble} Frontier Fields (HFF, GO: \#11507, PI: Lotz, \citealt{Lotz2017}) observations, which provided the deepest \hst observations of six massive strong-lensing clusters, including Abell\,2744 high precision strong-lensing analyses were performed \citep{Wang2015,Jauzac2015_a2744}. These were further complemented by high resolution spectroscopy with the Multi Unit Spectroscopic Explorer (MUSE) spectrograph at the Very Large Telescope (VLT) \citep{Mahler2018, Richard2021,Bergamini2023a}. \cite{Medezinski2016} performed a weak-lensing analysis, followed by a combined strong and weak-lensing analysis from \citet{Jauzac2016}. \citet{Jauzac2016} also detected the North and North-West substructures. The most recent lensing analyses of Abell\,2744 make use of \jwst, including strong-lensing \citep{Furtak2023b}, weak-lensing \citep{Harvey2024} and the combination of both strong and weak-lensing regimes \citep{Cha2024}.   

The work presented in this paper is built up on previous studies \citep{Jauzac2015_a2744,Jauzac2016,Mahler2018,Furtak2023b}, with a focus on achieving a more precise and ideally accurate modelling of the North and the North-West substructures of the cluster. These are two massive substructures that were identified in earlier analyses using \hst\ observations \citep{merten2011,Jauzac2016}, and \textit{JWST} observations revealed at least eight new multiple image systems \citep{Furtak2023b}, confirming their high density and the necessity to properly include them in the mass model of the overall cluster as independent substructures. 

This work makes use of the high resolution imaging provided by \hst\ from the Beyond Ultra-deep Frontier Fields And Legacy Observations \citep[BUFFALO, GO: \#15117, PIs: Steinhardt \& Jauzac,][]{Steinhardt2020} survey and the deep \jwst\ imaging from the Ultradeep NIRSpec and NIRCam ObserVations before the Epoch of Reionization \citep[][UNCOVER, GO: \#2561, PIs Labbe $\&$ Bezanson]{Bezanson2022} programme. BUFFALO is built upon the HFF by extending the area observed by the \hst\ for the six HFF clusters. The survey expanded the HFF footprint by approximately a factor of four, adding flanking fields around each cluster to reach a total area of $\sim$16 arcmin$^2$ per cluster \citep{Steinhardt2020}. The main goals of the survey are (1) to use gravitational lensing in massive clusters to study dark matter by combining deep strong-lensing and high resolution weak-lensing observations to improve the precision and accuracy of the clusters' mass models \citep{Gonzalez2020,Niemiec2023,Lagattuta2023,Perera2025}, and (2) to study the distant Universe using these strong lenses as cosmic telescopes \citep{Diego2024,pagul2024}. 
On the other hand, marking the beginning of the \textit{JWST} era in cluster lensing, the \textit{JWST} UNCOVER Treasury Survey is a Cycle 1 programme aimed at probing the early Universe by targeting Abell 2744 \citep{Furtak2023a,Furtak2023b}. Taking advantage of the superior sensitivity of \textit{JWST}, the survey combines ultradeep imaging and spectroscopy to detect and characterize galaxies formed before and during the Epoch of Reionization. BUFFALO's extended footprint provides wider field imaging that is essential to constrain the cluster mass distribution on larger scales and to improve the weak-lensing reconstruction in the outskirts which relied on ground-based observations until then. In parallel, the unprecedented depth and spatial resolution of UNCOVER reveal new multiple image systems, particularly in the North and North-West regions, where previous models were limited by the lack of strong-lensing constraints. The combination of these datasets therefore provides a major advance in both the strong and weak lensing regimes, enabling a more complete and better constrained description of the mass distribution in Abell\,2744. 

We here present a new strong and weak-lensing mass model of Abell\,2744 built by combining \hst\ and \jwst\ observations. The strong-lensing analysis is divided into three regions : the core, North and North-West. In the previous studies, due to the lack of constraints, the North and North-West regions were fixed while optimising the core. However, in this work, the new \jwst\ observations in these regions are used and the mass distribution in the cluster is optimised together for the core, North and North-West to obtain a high precision strong-lensing mass model of Abell\,2744. Similarly, for the weak-lensing analysis, the background galaxy catalogue is made by combining galaxies from \hst~and \jwst~selection. We use this mass model to identify substructures in Abell\,2744 and compare that with the BAryons and HAloes of MAssive Systems (BAHAMAS) simulations for CDM and different SIDM models. These simulations provide a statistically representative sample of massive galaxy clusters and allow us to investigate how the SHMF varies for different dark matter models, including CDM and SIDM. 

The paper is organised as follows: Section\,\ref{sec:obs} presents the data used for this work. Section\,\ref{sec: GL_products} presents the data products used for the gravitational lensing analysis. The strong and weak-lensing mass modelling of Abell\,2744 is described in Section\,\ref{sec:mass_modelling}. Section\,\ref{sec:bahamas_analysis} outlines the method used to construct the subhalo mass function in the BAHAMAS simulations. We discuss our results in Section\,\ref{sec:discussion}. Throughout the paper, we assume a flat cosmological model with $\Omega_M$ = 0.3, $\Omega_\Lambda$ = 0.7, and a Hubble constant H$_{0}=70$\,km\,s$^{-1}$Mpc$^{-1}$. In this cosmology, $1\arcsec = 4.7$\,kpc at the redshift of the cluster, $z = 0.308$. All magnitudes are given in the AB system \citep{Oke1974}.

\section{DATA}
\label{sec:obs}
As one of the most studied galaxy clusters, Abell\,2744 has been extensively targeted by numerous imaging and spectroscopic campaigns spanning multiple wavelengths. Deep \hst~imaging was obtained using the \hst~/Advanced Camera for Surveys (ACS) and \hst~/Wide Field Camera 3 (WFC3) as part of the HFF and the BUFFALO programmes, providing high resolution optical and near-infrared coverage. These are complemented by extremely deep \jwst~Near Infrared Spectrograph (NIRSpec) and Near Infrared Camera (NIRCam) observations from the UNCOVER programme, enabling detection of faint, high redshift galaxies. Spectroscopically, Abell\,2744 has been studied through the Grism Lens Amplified Survey from Space (GLASS; \citealt{Treu2015}), which combines HFF imaging with slitless spectroscopy, and follow up ground based spectroscopy from Keck and VLT/MUSE. Additionally, the cluster was observed in the submillimeter regime as part of the ALMA (Atacama Large Millimeter/submillimeter Array) Frontier Fields survey \citep{Gonzalez2017}. Recently, the MegaScience (PI: Suess) programme \citep{Suess2024}, a \textit{JWST}/NIRCam medium-band imaging survey, has provided further high quality photometric data. In the X-ray regime, observations from the \textit{Chandra X-ray Observatory} and the \textit{XMM-Newton Telescope}
have revealed complex thermal structures associated with the ongoing merger activity in the intracluster medium \citep[e.g.][]{Owers2011, Eckert2015, Chadayammuri2024, Finner2025}. This section only presents and discusses the observations used in this work. 

\subsection{\textit{Hubble Space Telescope} observations}
Abell\,2744 has been observed by \hst~through two major programmes. Initially, it was observed as part of the HFF programme between 2013 October 25 and 2014 July 1, using seven filters: three ACS filters (F435W, F606W, and F814W - 46ks, 24ks and 104ks respectively), and four WFC3 filters (F105W, F125W, F140W, and F160W - 68ks, 33ks, 28ks and 67ks respectively).
For this work, we utilise data from the BUFFALO programme, which observed Abell\,2744 between 2019 May 15 and 2019 December 7. BUFFALO observations used five filters: two ACS filters (F606W and F814W - exposure times of 1538s and 2934s respectively), and three WFC3 filters (F105W, F125W, and F160W - exposure times of 1426s, 1467s, and 1579s respectively). A list of all these observations is given in appendix in Tab.\,\ref{tab:obs}.

Individual exposures are used to facilitate PSF modelling for weak-lensing measurements. All \hst~data were astrometrically aligned relative to each other and registered to the Gaia absolute astrometric reference frame. The aligned data were then combined into mosaics with pixel scales of 30 mas for ACS and 60 mas for WFC3, following the methodology described in \citet{Koekmoer2011}.

\subsection{\textit{James Webb Space Telescope} observations}
Abell\,2744 was observed by \jwst~as a part of the UNCOVER survey. The cluster was observed in six broad-band filters, F115W, F150W, F200W, F277W, F356W and F444W, and in one medium band filter, F410M, for 4-6 hours each. The data were then reduced and drizzled into mosaics with the \texttt{grism redshift and line analysis software for space-based spectroscopy} \citep[\texttt{grizli,}][]{Brammer2021}. These mosaics were constructed by combining the \jwst~observations of Abell\,2744 from the GLASS-JWST survey \citep{Treu2022}, obtained with NIRCam and NIRISS, together with the NIRCam data acquired under the Director’s Discretionary Time programme (PID 2767; PI: Kelly). These publicly available NIRcam mosaic images of Abell\,2744 processed by the UNCOVER team are used in this work. With a field of view of $\sim$45 arcmin\textsuperscript{2}, the UNCOVER mosaics represents the widest ultra-deep imaging coverage of Abell\,2744 to date.  

\subsection{Spectroscopic Redshifts}
\label{sec:redshifts}
Spectroscopic coverage of Abell\,2744 is available throughout the \hst/BUFFALO footprint, drawn from a variety of publicly-available sources. The majority of redshifts come from a deep (19 hour) MUSE mosaic taken as part of the MUSE GTO lensing clusters programme (PI: Richard; \citealt{Mahler2018}.) This is supplemented by grism spectroscopy taken from the GLASS(PI: Treu; \citealt{Schmidt2014}) as well as other archival coverage available in the NASA Extragalactic Database (NED).

To build our spectroscopic redshift catalogue, we combine information from each individual source (RA, Dec, redshift, and confidence value) into a master list. We first visually inspect the positions of each entry by overlaying their coordinates on the BUFFALO images and adjusting the centroid values of each object to match the observed \hst. Next, we normalize the confidence values of each redshift to a common scale, choosing the 3 (secure redshift) / 2 (highly probable redshift) / 1 (low confidence/guess redshift) convention employed by the MUSE GTO team \citep[e.g.,][]{lagattuta2019,Richard2021}. Finally, we remove any duplicate entries that appear in multiple catalogues, leaving a single redshift entry for each distinct object. In cases where overlapping entries disagree on the final redshift, we select the entry with the higher confidence value. We obtain a final catalogue of 788 redshifts, which we then use for cluster members identification, foreground contamination removal, and lens modelling. 
Finally, we add the UNCOVER spectroscopic redshifts presented in \citet{Price2025}. Our final spectroscopic redshifts catalogue is composed of 1,170 sources.

\subsection{Photometric Redshifts}

In this work, photometric redshifts from the Medium Bands, Mega Science survey (MegaScience, PID 4111, PI: Suess) were used, one of the deepest multi-band photometric datasets currently available for high-redshift galaxy clusters \citep{Suess2024}. MegaScience is a \jwst~Cycle 2 survey executed in November 2023, built upon and extending the UNCOVER Cycle 1 Treasury survey. MegaScience provides a uniquely comprehensive dataset, combining deep ($\sim$28–30 mag$_{\mathrm{AB}}$) NIRCam imaging in eleven medium-band filters (F140M, F162M, F182M, F210M, F250M, F300M, F335M, F360M, F430M, F460M, and F480M) and the two shortest-wavelength broadband filters (F070W and F090W) over a $\sim$30\,arcmin$^2$ region centered on Abell\,2744. These data complement the UNCOVER observations, enabling precise spectral energy distribution (SED) fitting across the full suite of NIRCam filters. 
The catalogue is composed of 74,020 sources with photometric redshifts.\\

\subsection{BAHAMAS Simulations}
In order to interpret our observational measurements, we compare our results with theoretical predictions. For that purpose, we use the BAryons and HAloes of MAssive Systems \citep[BAHAMAS,][]{McCarthy2017} suite of hydrodynamical simulations.
BAHAMAS is designed to reproduce the global properties of massive halos, particularly galaxy clusters, by explicitly modelling the joint evolution of dark matter, gas, stars, and black holes. Its CDM runs are therefore an excellent benchmark for the expected substructure distribution in a $\Lambda$CDM Universe. However, to explore whether alternative dark matter physics could alter the abundance or radial distribution of subhalos, we also make use of the SIDM extensions of BAHAMAS developed by \citet{Robertson2019}. These simulations incorporate isotropic, velocity independent dark matter scattering with cross-sections of 0.1, 0.3 and 1.0 cm$^2$ g$^{-1}$, as well as a velocity dependent model in which interactions are suppressed at the high relative velocities typical of galaxy clusters.

\section{GRAVITATIONAL LENSING DATA PRODUCTS}
\label{sec: GL_products}

\subsection{Multiple image systems}
\label{sec:mult_images}
The positions, shapes, and geometries of multiple images from background galaxies are determined by the cluster mass distribution, source galaxy distances, and the alignment between observer, cluster, and sources. Therefore, we constrain Abell\,2744's mass distribution by using the positions of identified multiple image systems as constraints to our strong-lensing mass model of the cluster core.

Strong lensing features in the central region of Abell 2744 have been the subject of identification efforts spanning multiple observational campaigns. Prior to HFF observations, early lens models by \citet{merten2011}, \citet{Richard2014} and \citet{Johnson2014} constructed a catalogue of 55 multiple systems, including three secure spectroscopic redshifts \citep{Johnson2014}.
Subsequent analysis of the HFF data by \citet{Lam2014,Jauzac2015_a2744,Ishigaki2015} identified approximately 185 images. This work was complemented by spectroscopic redshifts of 7 lensed sources obtained by the GLASS team \citep{Wang2015}. The comprehensive HFF imaging, combined with GLASS and MUSE spectroscopic observations, ultimately identified 60 multiply-imaged galaxies in Abell\,2744, generating 188 individual images throughout the cluster core \citep{Mahler2018,Richard2021}.
To ensure maximum reliability, this catalogue was subsequently refined by \citet{Bergamini2023a} to create a secure, purely spectroscopic sample. The strong-lensing model presented here employs spectroscopically confirmed images only, which comprises 90 multiple images from 30 distinct background sources, to accurately constrain the mass distribution within the main cluster core.

In addition to spectroscopically confirmed multiple images in the core, we use the latest \jwst\ multiple images identification in the North (N) and North-West (NW) regions of the cluster \citep{Furtak2023a,Price2025} to constrain these outer regions. Pre-\jwst\ studies had no spectroscopically confirmed multiple systems in the N region, and only one system \citep{Bergamini2023a} in the NW region.
With \jwst observations from UNCOVER, one spectroscopically confirmed system is now identified in the N region, comprising three multiple images that are further resolved into two distinct clumps each. In the NW region of the cluster, seven spectroscopically confirmed systems with 22 multiple images have been identified \citep{Furtak2023a,Price2025} and are included into the strong-lensing analysis in this work. A list of all 101 spectroscopically confirmed multiple images (from 37 systems) used in our model is given in Table\,\ref{tab:a2744_arcs} (in appendix).

\subsection{Cluster Members}
\label{sec:cluster_members}
The mass modelling technique used in this paper accounts for the matter contribution from cluster members. Consequently, a well defined catalogue of cluster members is needed. We here present the process to get such a catalogue.

\begin{figure}
    \centering
    \includegraphics[width=1\linewidth]{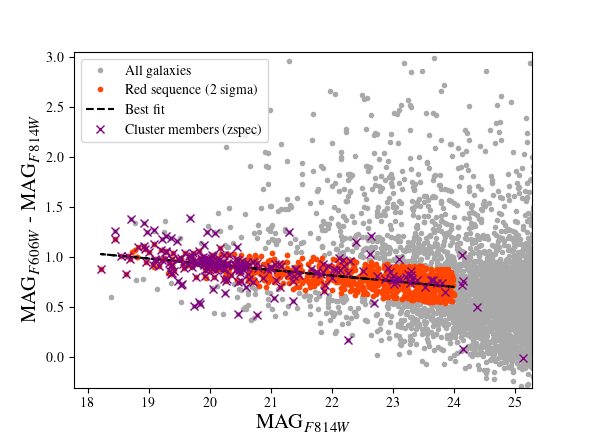}
    \caption{Colour (MAG$_{\rm F606W}$ - MAG$_{\rm F814W}$) vs magnitude (MAG$_{\rm F814W}$) diagram for Abell\,2744. All sources from the source extraction done in this work are shown in grey, and red sequence galaxies are highlighted in red. Cluster members with spectroscopic redshifts are marked as purple crosses. We consider all galaxies within the 2\,$\sigma$ of the linear fit as cluster members.}
    \label{fig:a2744_RS}
\end{figure}

We first used the spectroscopic redshift catalogue described in Section\,\ref{sec:redshifts} to compile a catalogue of secured cluster members.
Galaxies at the cluster redshift, i.e., 0.29 $< z <$ 0.32, are considered as cluster members and form the foundation to define the red sequence \citep{Gladders2000}. A colour-magnitude diagram ($m_{\rm F606W} - m_{\rm F814W}$ vs $m_{F814W}$) is constructed using spectroscopically identified cluster galaxies within this redshift range. For galaxies without spectroscopic redshifts, \textsc{SExtractor} is used in dual mode to extract \hst/F606W and \hst/F814W photometry. These photometrically selected candidates are then located on the same colour-magnitude parameter space to identify red sequence cluster members, as shown in Fig.\,\ref{fig:a2744_RS}. The red sequence is then fitted with a linear relationship using an iterative 3$\sigma$ clipping process. This iterative approach eliminates blue cluster member galaxies and defines spectroscopically confirmed red sequence while its scatter is measured. The red sequence is defined as galaxies falling within a region in colour-magnitude space bounded by 2$\sigma$ above and below the linear fit, with the bright limit set by the BCG (Brightest Cluster Galaxy) \hst/F814W magnitude and the faint limit set to 24 mag. To note, spectroscopically confirmed cluster galaxies fainter than 24 mag, are retained. From this catalogue, only the galaxies within 10\arcsec of a multiple image are used in our model. This is to capture the dominant deflectors (brighter galaxies are more massive) and local perturbations that most strongly influence the lensing, while the number of free parameters is dramatically reduced, making the optimisation computationally feasible. The final catalogue contains 324 galaxies including the five BCGs - two in the core, two in the NW and one in the N structures. 

\subsection{\hst~Weak lensing catalogue}
\label{sec:hst_wlcat}
In the weak lensing regime, the shapes of lensed background sources carry the lensing signal, and, as a result, are used to constrain the mass distribution. Such signal is weak and thus indistinguishable from the intrinsic shape for a single background source. Therefore, a statistical analysis is necessary. The construction of the weak-lensing background galaxy catalogue is presented in this section. 

\subsubsection{The ACS source catalogue}
We use the \textit{HST}/ACS F814W filter to detect sources and measure their shapes. Shape measurements are performed using \textsc{pyRRG},\footnote{https://pypi.org/project/pyRRG/} a publicly available weak-lensing shape measurement algorithm based on \cite{Rhodes2000} and presented in \cite{Harvey2019}.
Source extraction is performed using \textsc{Source-Extractor} \citep{Bertin1996} via the `hot-cold' method of \citet{Rix2004}. This method involves two \textsc{SExtractor} scans with different minimum thresholds: a hot scan to identify small and faint sources, and a cold scan to detect large and bright sources. The resulting catalogues are then combined, with the method ensuring that larger objects are retained.
This combined catalogue undergoes cleaning to remove duplicate detections and sources near stars or saturated pixels. Sources are then classified as stars, galaxies, or fake detections based on their distribution in the magnitude (MAG$\_$AUTO) vs peak surface brightness (MU$\_$MAX) parameter space \citep[see][for more details]{Leauthaud2007,jauzac2014,Richard2014}. Fake detections are removed from the catalogue, while the stellar sequence is retained for point spread function (PSF) modelling, a mandatory step to measure galaxy shapes.

\subsubsection{Shape measurements}
The next step involves measuring the shapes of background objects. Since instrumental effects and telescope positioning impact the true shapes of galaxies, these effects must be properly accounted for. \textsc{pyRRG} was constructed to correct for such effects by modelling the PSF of individual exposures for all images considered and correcting all galaxy shapes accordingly.
PSF modelling is performed by measuring the PSF from stars in each individual exposure, comparing their second and fourth order moments to various \textsc{Tiny Tim} models \citep{Krist2011}. \textsc{Tiny Tim} is a standard \textit{HST} modelling software that uses detailed information on the telescope and instruments to model the PSF of different instruments under various observational conditions. \textsc{pyRRG} combines the best-fit PSF over all individual exposures to produce a final PSF model that matches the observations at hand. With the modelled PSF, galaxy moments are then corrected and galaxy shapes are measured.

\subsubsection{Background galaxy selection}
\begin{figure}
    \centering
    \hspace{-1.7cm}
    \includegraphics[width=1.1\linewidth]{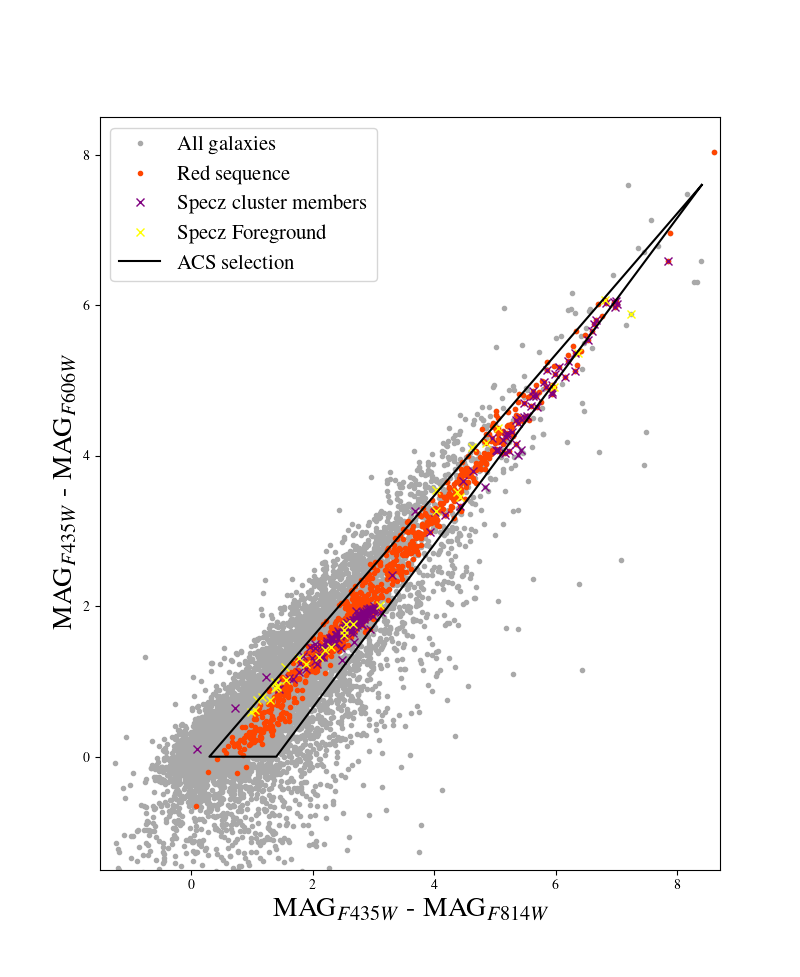}
    \caption{A colour-colour diagram showing the selection of background galaxies (outside the triangular region) for Abell\,2744. Cluster members identified from the red sequence selection are shown in red, and spectroscopically confirmed cluster members are shown in purple. Foreground galaxies with spectroscopic redshifts are shown in yellow. The solid black line denotes the ACS selection to separate the background galaxies from the foreground and cluster populations}.
    \label{fig:a2744_BG}
\end{figure}

At this stage, we have a weak-lensing catalogue containing galaxy shape measurements obtained with \textsc{pyRRG}. Nevertheless, foreground and cluster galaxies still contaminate this catalogue. While background galaxies carry the weak-lensing signal, their contribution is diluted by these contaminants. 
To obtain a cleaner catalogue, we exclude confirmed cluster member galaxies identified via the red-sequence selection. Even after removing these cluster members, contamination persists from faint, compact, and bluer cluster galaxies (which mimic background object colours), as well as foreground sources lacking redshift measurements. Therefore, we apply a colour-colour selection criterion for galaxies without available photometric and/or spectroscopic redshifts \citep{Jauzac2012,Jauzac2015_m0416,Jauzac2016,Medezinski2016,Medezinski2018,Niemiec2023}. Galaxies with photometric or spectroscopic redshifts are used to calibrate regions dominated by this unlensed population (cluster and foreground galaxies). Weakly lensed galaxies are then separated from the unlensed population by a defined polynomial region in the colour-colour parameter space, i.e., (MAG$_{\rm F435W}$ - MAG$_\mathrm{F814W}$) $vs$ (MAG$_\mathrm{F435W}$ - MAG$_\mathrm{F606W}$), as shown in Fig.\,\ref{fig:a2744_BG}. The sources outside the triangular region marked by solid black line are the selected background galaxies.

To further clean the catalogue, additional quality cuts are applied: (i) sources located close to the edges of different exposures are removed by applying a cut on the number of exposures for each detected source as done in \citet[][N23, hereafter]{Niemiec2023}, (ii) only sources with ellipticity in the range $0 < e < 1$ are retained, (iii) sources with signal-to-noise ratio $S/N < 4.5$ are removed, (iv) only sources with size parameters in the range $3.6 < d < 30$ pixels are kept following \citet{Jauzac2012,Jauzac2016,Niemiec2023}, (v) very faint sources are removed with an F814W magnitude cut at 29 in the HFF footprint and 27 elsewhere following \citetalias{Niemiec2023}, and finally (vi) very bright sources with F814W magnitude brighter than 23 are removed across the entire BUFFALO footprint. We remove all sources within the strong-lensing regions (i.e., the core, N, and NW regions), as these regions are not within the weak-lensing regime. This treatment also ensures that strongly lensed multiple images are excluded from our weak-lensing catalogue. The final catalogue contains 3308 background galaxies suitable for the weak-lensing analysis. This corresponds to a source density of $\sim$57\,galaxies.arcmin\textsuperscript{-2}, compared to other \hst~analysis, such as $\sim$110\,galaxies.arcmin\textsuperscript{-2} from \citet{Jauzac2016}, $\sim$42\,galaxies.arcmin\textsuperscript{-2} from \citetalias{Niemiec2023} in HFF, and $\sim$13\,galaxies.acrmin\textsuperscript{-2} in the Subaru sample and $\sim$25\,galaxies.acrmin\textsuperscript{-2} from \citet{Finner2025} with Subaru.

\subsection{\jwst~Weak lensing catalogue}
\label{sec:jwst_wlcat}
For the \jwst~weak-lensing catalogue, UNCOVER imaging in the F115W, F150W, and F200W filters are used. Following the same approach as for \hst, \pyrrg~\citep{Harvey2019,Harvey2024} is used for source detection, photometry, and galaxy shape measurements.

Spectroscopic redshifts from UNCOVER and photometric redshifts from MegaScience are used to retain only sources with $z > 0.32$, ensuring the selection of background galaxies behind the cluster. Additional quality cuts are applied to remove sources with ill-defined shapes, including signal-to-noise, ellipticity, and size cuts implemented in the same manner as described before for \hst~data. To preserve the quality of the lensing signal, sources with photometric redshift uncertainties exceeding 1$\sigma$ are removed. Similarly, very bright (mag $< 20$) and very faint (mag $> 27.5$) sources are excluded for the same reason, as done in \citet{Harvey2024}.
To further refine the catalogue and ensure contamination-free selection, cross-matching is performed with cluster members and foreground galaxies identified from BUFFALO, with any matches subsequently removed. A comparison of shear measurements across the F115W, F150W, and F200W filters revealed that the weak-lensing signal is optimal in F150W \citep{Harvey2024}. Therefore, this analysis uses weak-lensing measurements derived from the F150W filter.

The final \jwst~weak-lensing catalogue contains 2353 sources, corresponding to a source density of $\sim$50\,galaxies.arcmin$^{-2}$, compared to $\sim$150\,galaxies.arcmin$^{-2}$ from \citet{Harvey2024} and $\sim$350\,galaxies.arcmin$^{-2}$ \citep{Cha2024}.

\section{GRAVITATIONAL LENSING ANALYSIS}
\label{sec:mass_modelling}
\subsection{Strong Lensing Mass Modelling}
\label{sec:sl}
In the cluster core, where we assume light traces mass, i.e. that dark matter halos (at both cluster and galaxy-scales) follow the light distribution, we use a parametric mass model. The total mass distribution in this region is decomposed into large and small scale components to model cluster and galaxy-scale matter distributions respectively. Each component is modelled assuming a dual Pseudo Isothermal Elliptical density profile \citep[dPIE,][]{Eliasdottir2007} 
characterised by seven parameters: position ($\alpha$, $\delta$), position angle ($\theta$), ellipticity (e), velocity dispersion ($\sigma$), and two scale radii, the core and the cut radii (\textit{r}$_{\rm core}$ and \textit{r}$_{\rm cut}$). The 3D density distribution of the dPIE is written as 
\begin{equation}
    \rho(r) = \frac{\rho_{0}}{(1 + r^{2}/r^{2}_{\rm{core}})(1 + r^{2}/r^{2}_{\rm{cut}})}, \,\,\, \rm{with}\, \textit{r}_{\rm{cut}} > \textit{r}_{\rm{core}}.
\label{eq:dpie}
\end{equation}
where $\rho_0$ is the central density of the core. To reduce the number of free parameters in the lens model, position, ellipticity and angle position of galaxy-scale components are fixed to follow their light distribution. In addition, the Faber-Jackson empirical scaling relation \citep{Faber1976} are used for the remaining parameters. In this case, we consider that for each cluster galaxy, the core radius, \textit{r}$_{\rm core}$, the cut radius, \textit{r}$_{\rm cut}$ and the velocity dispersion, $\sigma_0$, scale with the galaxy luminosity, L, as follows:
\begin{equation}
\label{eq:scaling_relation}
\begin{split}  
\sigma_{0} & = \sigma_{0}^{\star} \left(\frac{L}{L_{\star}}\right)^{1/4}, \\
r_{\rm core} & = r_{\rm core}^{\star} \left(\frac{L}{L_{\star}}\right)^{1/2}, \\
r_{\rm cut} & = r_{\rm cut}^{\star} \left(\frac{L}{L_{\star}}\right)^{\alpha}
\end{split}
\end{equation}
where L\textsuperscript{$\star$} is the typical luminosity of an elliptical galaxy at the cluster redshift, and \textit{r}$_{\rm core}^{\star}$, \textit{r}$_{\rm cut}^{\star}$, $\sigma_0^{\star}$ are the dPIE parameters. $\alpha$ is the slope which is fixed to 1/2, following the work by \cite{Niemiec2020,jullo2009}.

The publicly available lens modelling software \textsc{Lenstool} is used for the mass reconstruction. Unlike previous studies where parametric modelling was performed by optimizing the core while keeping the N-NW regions fixed, here, all three regions are optimised simultaneously. 

The core contains two cluster-scale halos, while the N and NW regions contain three cluster-scale halos, one in the N and two in NW region. All parameters for these halos are set to vary freely except for the cut radius, which generally lies outside the strong-lensing region, making it difficult to be constrained by our model. It is thus fixed to \textit{r}$_{\mathrm{cut}}$=1000\,kpc. The cluster contains two BCGs in the core, two in the NW region, and one in the N region. All cluster-scale halos are initially centred on the BCGs in their respective regions. The positions of the core (DM1, DM2) and north-west (DM3, DM4) halos are optimised, while the position of the northern halo (DM5) is held fixed at its BCG, since the single spectroscopically confirmed multiple-image system in the N region does not sufficiently constrain the halo position.

For small-scale halos, cluster members are modelled using scaling relations with $mag_0$ = 19.44 at the Abell\,2744 redshift in the ACS/F814W passband. Their positions are fixed to those of the corresponding cluster galaxies, along with their position angles and ellipticities. Core radii are fixed to the small value of $r_{\rm core}=0.15$\,kpc while cut radii and velocity dispersions are set to vary freely with priors derived from previous models. The two BCGs in the core are modelled separately from the other cluster members. 

The final model, where all cluster components are modelled simultaneously with 34 free parameters, yields an rms of 0.83\arcsec. The parameters of this strong-lensing mass model that minimises the $\chi^2$ between the observed and model predicted positions of the multiple image systems, the best-fit parameters, are listed in Table\,\ref{tab:a2744_sl}. The corresponding surface mass density of Abell\,2744 as a function of distance from the cluster center is shown in yellow in Fig.\,\ref{fig:a2744_sdp1}. From the best-fit model, we measure a projected mass of \textit{M}(${R}<200{\mathrm{kpc}})=(1.69\pm 0.03)\times$10$^{14}$\,M$_{\mathrm\odot}$. 

\begin{figure*}
  \centering
  \subfloat[]{
    \includegraphics[width=0.48\linewidth]{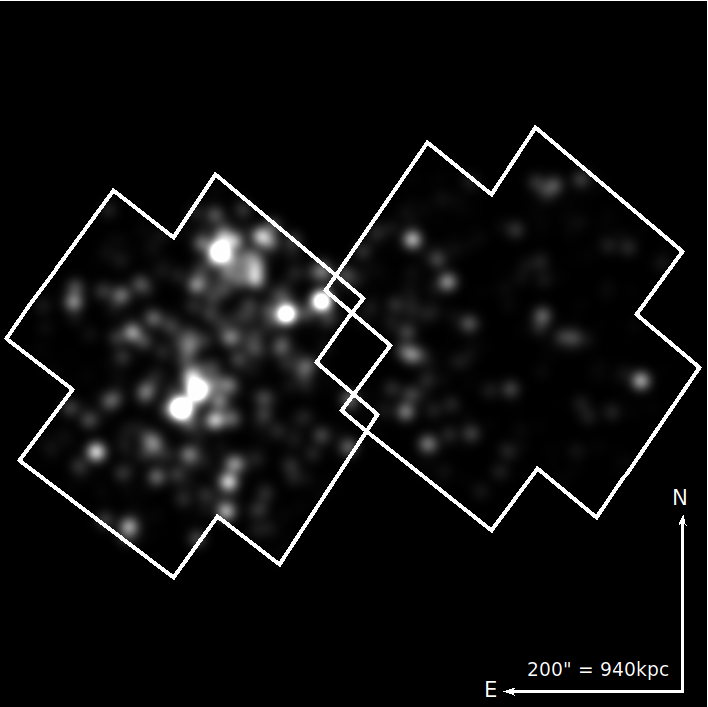}
    \label{fig:lightmap}
  }
  \hfill
  \hspace{-0.5cm}
  \subfloat[]{
    \includegraphics[width=0.48\linewidth]{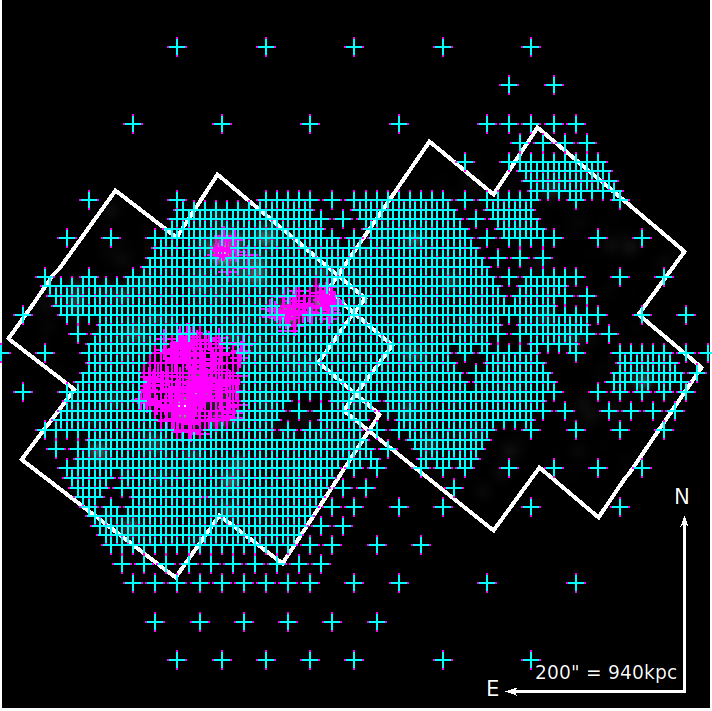}
    \label{fig:a2744_grid}
  }
  \caption{Left: Smoothed lightmap of Abell\,2744. Right: Locations of the non-parametric grid potentials used in the weak-lensing-only and combined strong+weak lensing analyses, shown by magenta and cyan crosses, respectively. Crosses mark the centres of the grid potentials. In the combined model, the grid potentials in the core, N and NW regions are removed to avoid overlap with the strong-lensing constraints. Since the remaining grid potentials are common to both models, the cyan crosses obscure the overlapping magenta crosses in these regions. The white solid line outlines the BUFFALO field of view.}
  \label{fig:a2744_lightmap_grid}
\end{figure*}

\subsection{Weak Lensing Mass Modelling}
\label{sec:wl}
Unlike the parametric strong-lensing analysis used for cluster cores, our approach utilises a non-parametric mass modelling methodology to reconstruct the mass distribution in the cluster outskirts, where the weak-lensing signal is predominant
\citep{Jauzac2012,Jauzac2015_m0416,Jauzac2016,Jauzac2018,Gonzalez2020,Niemiec2020,Niemiec2023}.
Our weak-lensing analysis is based on a multi-scale potential grid to decompose the matter distribution. This multi-scale grid technique \citep{jullo2009,Jauzac2012} utilises an irregular grid consisting of Radial Basis Functions (RBFs) to enhance the flexibility of the lensing mass reconstruction. These RBFs are radially symmetric real valued functions positioned at each grid node. The multi-scale grid is constructed from a smoothed cluster lightmap and undergoes recursive refinement in regions of highest density. A lightmap is generated using the spatial distribution of cluster member galaxies. We employ the \textsc{SExtractor} segmentation map to isolate the light from these galaxies in the \textit{HST} imaging, which is then convolved with a Gaussian kernel to produce the smoothed lightmap. A hexagonal grid geometry centered on the cluster core is applied to this smoothed lightmap, with recursive subdivision into six equilateral triangles at each refinement level. Each RBF is parameterised using a truncated isothermal mass distribution (TIMD profile, representing a circularly symmetric version of the dPIE profile). This profile comprises a two component pseudo isothermal mass distribution characterised by a core radius, s, set equal to the separation between an RBF and its nearest neighbour, and a cut radius, t \citep{jullo2009}. The TIMD profile includes a scaling factor, $v_{i}^2$, which represents the central velocity dispersion of the gravitational potential, $i$. For all grid potentials, the cut radii are set to a fixed ratio of $t=3s$ \citep{jullo2009}.

We combine the two catalogues from BUFFALO and UNCOVER. For sources present in both catalogues, those from UNCOVER are retained due to \jwst's superior spatial resolution and lower noise characteristics, which provide more accurate shape measurements and improved weak-lensing signal quality. The combined catalogue contains 5209 galaxies, corresponding to a source density of $\sim$66 galaxies per arcmin$^2$.

A multiscale grid is built using a lightmap to decompose the mass distribution. After testing different grid resolutions, we converge on an optimal grid composed of 1392 RBFs \citep{Jauzac2012,Jauzac2015_m0416,Niemiec2023} whose amplitudes constitute the 1392 free parameters of the weak-lensing model. Core radii, \textit{s}, are ranging from 14\arcsec\ to 171\arcsec. The multi-scale nature of the RBF grid implies that the effective spatial resolution of the reconstructed mass map varies across the field, as the grid refinement follows the smoothed cluster light distribution. Consequently, the reconstruction of substructures may retain some dependence on the local RBF scale, particularly for structures approaching the effective resolution of the grid. This should be considered when interpreting the properties and masses of individual substructures identified from the reconstructed mass map. The resulting grid is shown in Fig.\,\ref{fig:a2744_grid} on the smoothed cluster light map. The surface mass density for the corresponding model is shown in Fig.\,\ref{fig:a2744_sdp1}. The projected surface mass within 1000\,kpc, \textit{M}(${R}<1000\,{\mathrm{ kpc}})=(1.32\pm 0.11)\times$10$^{15}$\,M$_{\mathrm\odot}$.

\subsection{Combined Strong and Weak Lensing Mass Modelling}
\label{sec:sl+wl}
\begin{figure*}
  \centering
  \subfloat[]{
    \includegraphics[width=0.48\linewidth]{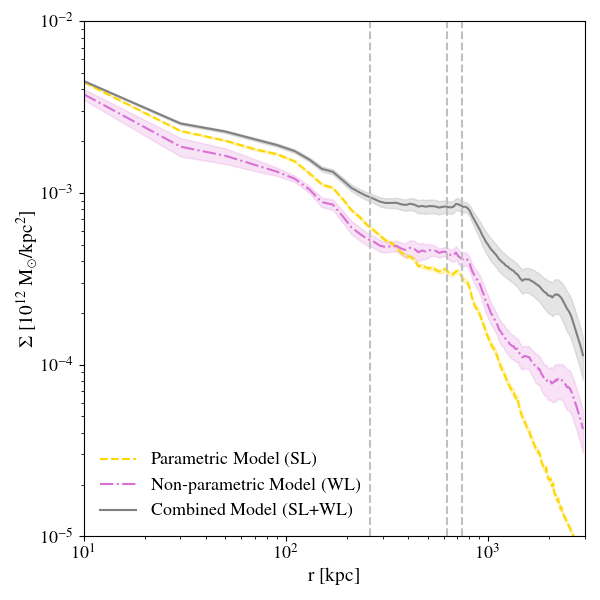}
    \label{fig:a2744_sdp1}
  }
  \hfill
  \subfloat[]{
    \includegraphics[width=0.48\linewidth]{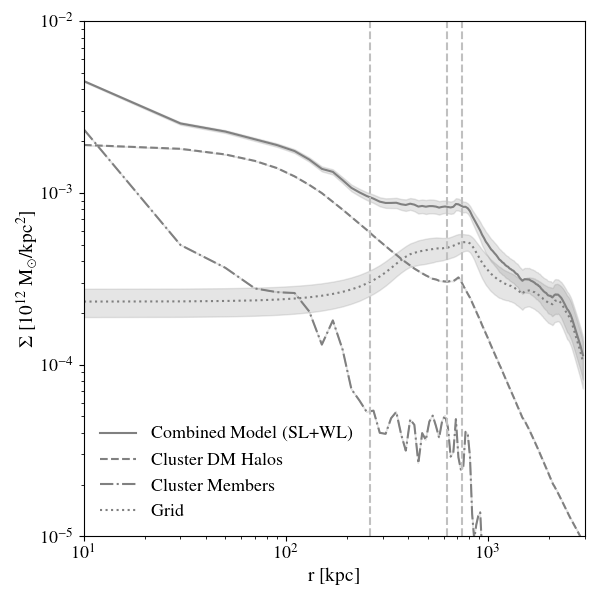}
    \label{fig:a2744_sdp2}
  }
  \caption{Surface mass density profiles of Abell\,2744. 
  \textit{Left:} Comparison of parametric strong lensing only (yellow), non-parametric weak-lensing only grid model (magenta), and combined strong+weak-lensing analysis (grey). 
  \textit{Right:} Decomposition of the full combined strong and weak-lensing model into cluster-scale halos (Core, N, NW: dashed), cluster members (dotted), and grid potentials (dash–dotted). 
  The vertical grey dashed lines mark the core radius at 260\,kpc and the NW and N substructures at 620\,kpc and 740\,kpc, respectively.}
  \label{fig:a2744_sdp}
\end{figure*}

To obtain a full, strong and weak-lensing model of the cluster, a sequential fit is performed by combining the parametric and non-parametric models. First, the core is modelled using strong-lensing constraints as described in Sect.\,\ref{sec:sl}. To the best-fit parametric model of the core, the multi-scale grid of potentials described in Sect.\,\ref{sec:wl} is added to model the outskirts of the cluster. The parametric model presented in Sect.\,\ref{sec:sl} is fixed to its best-fit values. RBF amplitudes are then estimated from the weak-lensing constraints. Already optimised cluster-scale dark matter halo and cluster member galaxies are combined with RBF grid potentials. The grid potentials are removed from the cluster core and N and NW substructures due to the absence of weak-lensing constraints in these regions. After this removal, 1300 grid potentials remain, whose amplitudes constitute the 1300 free parameters of the combined model. We note that a simultaneous strong+weak-lensing fit is in principle preferable, but requires further validation on a cluster as dynamically complex as Abell\,2744 before being applied here. Our team is currently working on developing and testing such an approach.
The projected surface mass density resulting from the combined strong and weak-lensing mass model of Abell\,2744 is shown in Fig.\,\ref{fig:a2744_sdp1} and Fig.\,\ref{fig:a2744_sdp2}. A total (combined strong and weak lensing) projected mass of \textit{M}($R<1000\,{\mathrm{kpc}})=(2.50\pm 0.20)\times 10^{15}$\,M$_{\mathrm{\odot}}$ is measured within 1000\,kpc from the combined strong and weak-lensing analysis.

\begin{figure*}
    \centering
    \includegraphics[width=1.\linewidth]{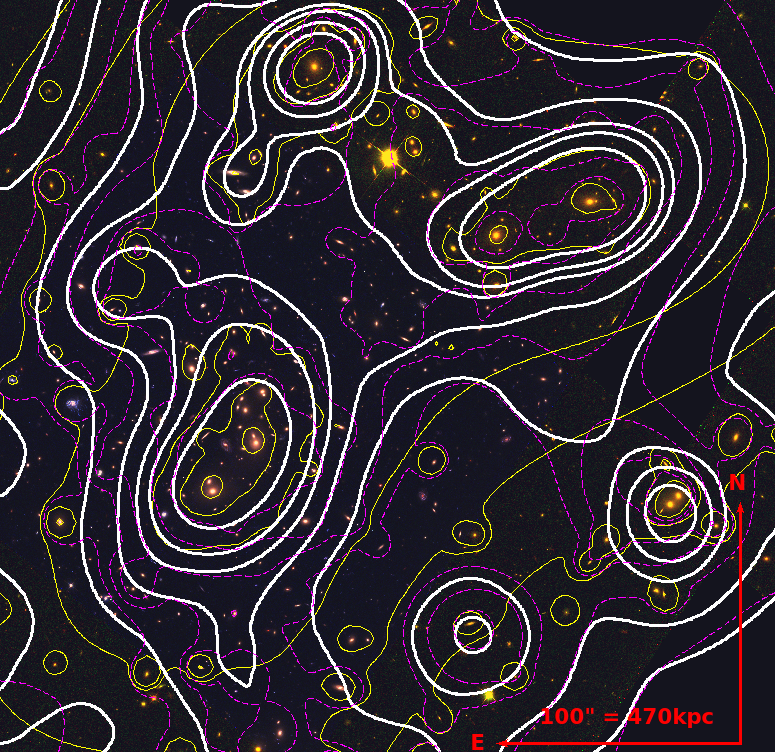}
    \caption{\textit{HST} colour composite image of the central region of Abell\,2744 created using the F435W for blue, F606W for green and F814W for red colours. The thin solid yellow and thin dashed magenta contours show the mass distribution obtained with our strong-lensing and weak-lensing mass reconstruction as described in Sect.\,\ref{sec:sl} and Sect.\,\ref{sec:wl} respectively. The mass contours obtained with our combined mass model, as presented in Sect.\,\ref{sec:sl+wl}, are shown with thick solid white contours.}
    \label{fig:a2744_contours}
\end{figure*}

\subsection{Substructure Detection}
\label{sec:substructure_detection}

The presence of substructures can be inferred from the projected density profiles where the peaks in the profiles are caused by overdensities in the mass maps. However, in order to detect the substructures from the mass map of Abell\,2744 in a more systematic manner, a wavelet transform (WT) is used to identify the positions of the peaks on sub-cluster scales in the mass map as done in \citet{schwinn2018}. The WT has been widely adopted in signal and image processing due to its ability to isolate features at specific scales while retaining spatial information. It is a mathematical tool that allows a signal to be analysed in both position and scale simultaneously. In this framework, the signal is decomposed into a series of components by convolving it with a set of basis functions, called wavelets, that are derived by scaling and translating a common reference known as the mother wavelet. This process enables the detection of localised features at specific spatial scales, in contrast to traditional techniques that separate signals purely by frequency. As a result, structures of different sizes can be isolated while preserving information about their exact locations.

When applied to two-dimensional mass maps of galaxy clusters, the wavelet transform allows the extraction of substructure signals at cluster relevant scales. The transformation produces a set of wavelet coefficients, each of which quantifies the strength of the signal at a given spatial position and scale. A high wavelet coefficient at a particular location implies a strong resemblance between the local mass distribution and the wavelet shape at that scale, indicating the presence of a significant feature, such as a mass peak. Conversely, low coefficients correspond to either noise or smooth background structure. 

In this work, the wavelet transform is used to identify significant substructures within the projected mass maps. The wavelet approach has been used successfully in several previous studies of cluster substructure, demonstrating its utility for identifying significant structures across a range of cluster mass maps \citep{Flin2006,Livermore2017,schwinn2018}. A set of selection criteria is defined based on both the spatial scale and the amplitude of the wavelet coefficients. Only peaks exceeding a threshold calibrated using the Abell\,2744 mass map are retained. When comparing with simulations, the same wavelet based detection procedure and selection criteria are applied to the projected mass maps, providing a consistent two-dimensional identification of substructures across the observational and simulated datasets. The mass maps themselves, however, are constructed differently: the Abell 2744 map is reconstructed from gravitational lensing constraints using a multi-scale RBF grid informed by the observed light distribution, whereas the simulated maps are constructed directly from projected particle masses. The resulting comparison should therefore be interpreted with this difference in the underlying mass-map reconstruction in mind.

The one-dimensional (1D) continuous wavelet transform is performed using the publicly available Python package \texttt{pywt} \citep{Lee2019}. The Mexican Hat wavelet is chosen as the mother wavelet because its shape closely resembles the expected density profile of subhalos and provides the best match compared to alternative wavelet functions. Although the input is a two-dimensional (2D) mass map, the wavelet transform is applied in a separable manner: first along the rows and then along the columns of the map. At each position, the wavelet coefficients quantify the strength of the local signal at a given spatial scale. The final coefficient map is obtained by computing the arithmetic mean of the row-wise and column-wise coefficients, yielding a scale sensitive and spatially resolved representation of potential substructures. To detect significant substructures, a quantitative selection criterion is defined based on two parameters: the scale of the wavelet transform and a threshold on the coefficient amplitude. These parameters are calibrated to maximise recovery of substructures identified in the density profiles from the lensing analyses in this study and by \citet[][J16, hereafter]{Jauzac2016}, resulting in the adoption of a WT scale of 40 pixels (182\,kpc) and a coefficient threshold of W $\geq$ 2.6 $\times$ 10\textsuperscript{10} M$_\odot$ pc\textsuperscript{-1}. 

All pixels with coefficients at least five times greater than the average value of the map are initially flagged as potential substructure locations. Since the wavelet convolution produces extended peaks, a single physical substructure is typically represented by many neighbouring pixels above this threshold. Rather than processing every flagged pixel, 20$\%$ of the pixels are randomly sampled as initial seed positions, reducing computational overhead while retaining sufficient coverage of all candidate peaks. Around each of the selected pixels, a circular aperture of radius 100\,kpc is drawn. The pixel with the highest wavelet coefficient within each aperture is identified, and the aperture is recentered on that pixel. This process is iterated ten times, allowing the seed positions to converge to the local maxima of the wavelet coefficient map. This iterative recentering ensures robust detection of all statistically significant peaks while enforcing a minimum spacing of 100\,kpc between substructures. This distance criterion avoids overlapping apertures, which would hinder independent mass estimates. Lastly, any substructure with an aperture mass M($<$150\,kpc)$<3 \times 10^{13} M_\odot$, or located further than 2.5\,Mpc from the cluster centre (end of the parallel field), is discarded from the final sample.

\begin{figure*}
    \centering
    \includegraphics[width=\textwidth]{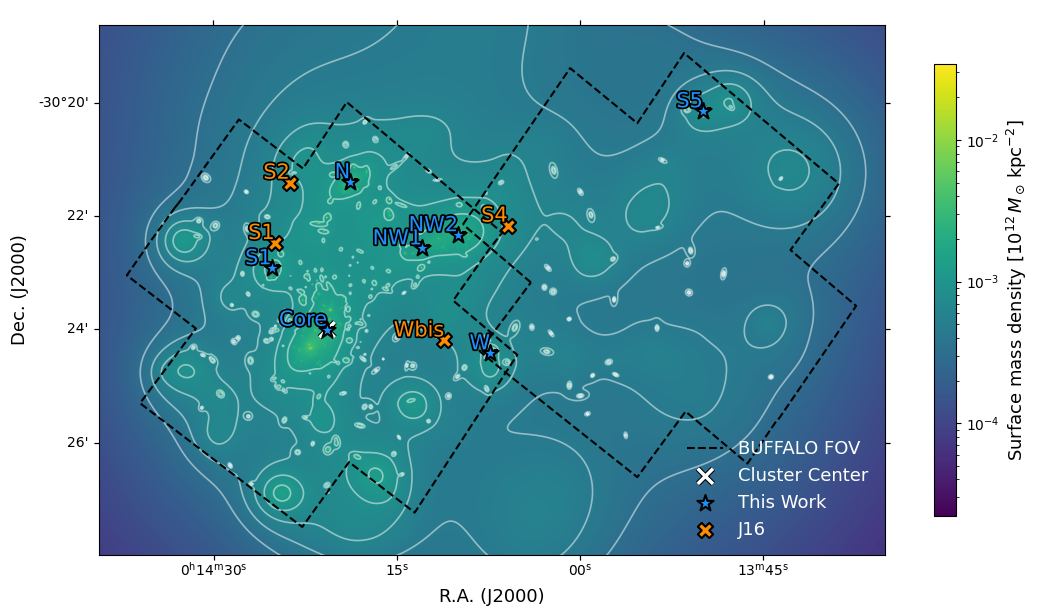}
    \caption{Substructures in Abell\,2744 identified using the wavelet transform method in the strong+weak-lensing mass map. The colour map and contours show the 2D surface density. Substructures detected in this work are marked in blue coloured stars while the ones from \citetalias{Jauzac2016} are highlighted as orange crosses. The BUFFALO field of view is shown by black dashed line.}
    \label{fig:WT_subs}

    \vspace{0.5cm}

    \begin{minipage}{\textwidth}
    \centering
    \captionof{table}{Substructure detections with the WT method. This work detects seven substructures, Core, N, NW (NW1 and NW2), W, S1 and S5. First column gives the ID of these substructures, their R.A. and Dec. (J2000) are given in column 2 and 3 respectively. Column 4 gives the mass enclosed within 150\,kpc (in 10$^{13}$M$_\odot$) and their distance to the cluster center (D$_{C-S}$) is listed in column 5. Column 6 compares the mass (within 150\,kpc) to that of \citetalias{Jauzac2016}.}
    \label{tab:a2744_subs}
    \begin{tabular}{c c c c c|c}
        \hline\hline
         ID & R.A. & Dec. & M(r<150\,kpc) & D$_{C-S}$ & M$_{\mathrm{J16}}$(r < 150\,kpc) \\ 
         & (deg) & (deg) & (10$^{13}$M$_\odot$) & (kpc) & (10$^{13}$M$_\odot$) \\
         \hline 
         \textit{Core} & 3.586484 & -30.400171 & 13.14 $\pm$ 0.28 &  -  &  13.55 $\pm$ 0.09 \\
         \textit{N}    & 3.578403 & -30.356798 & 9.56  $\pm$ 0.46 & 739 &  6.10 $\pm$ 0.50  \\
         \textit{NW1}  & 3.553918 & -30.376269 & 9.68  $\pm$ 0.37 & 619 &  7.90 $\pm$ 0.60 \\
         \textit{NW2}  & 3.541670 & -30.372377 & 10.53 $\pm$ 0.31 & 799 &  6.50 $\pm$ 0.60 \\
         \textit{W}    & 3.530713 & -30.407135 & 5.89  $\pm$ 0.46 & 815 &  - \\
         S1            & 3.605168 & -30.382088 & 7.22  $\pm$ 0.51 & 410 &  5.00 $\pm$ 0.40 \\
         S5            & 3.458225 & -30.335933 & 8.42  $\pm$ 0.41 & 2152& - \\
         \hline\hline
    \end{tabular}
    \end{minipage}
\end{figure*}

This method successfully detects the core, N and NW structures in the cluster. Apart from this, it also finds comparatively less massive substructures in the West of the cluster, W, and the East, S1, as shown Fig.\,\ref{fig:WT_subs}. There is also another detection, S5, in the parallel field of BUFFALO, around 2.1\,Mpc away from the cluster center. This detection has a low signal to noise, and as a result is considered as a candidate. This is the first detection of S5, thanks to the wide field of view of BUFFALO. It is associated to at least ten cluster members identified from the red sequence but the lack of spectroscopic redshifts in this region does not allow us to confirm it. 

\citetalias{Jauzac2016} detected eight substructures, core, N, NW (S3 and S4), West (Wbis), S4 which is located near the NW substructure, and two more S1 and S2 in the East, as shown in Fig.\,\ref{fig:WT_subs}. Table\,\ref{tab:a2744_subs} shows comparison of masses of the detected substructures.
The core component of the cluster has a mass of $M_{\mathrm{SL+WL_{core}}}(R < 150\,\mathrm{kpc}) = (13.14 \pm 0.28) \times 10^{13}$ M$_\odot$, which is in agreement with \citetalias{Jauzac2016}, $M_{\mathrm{J16_{core}}}(R < 150\,\mathrm{kpc}) = (13.55 \pm 0.09) \times 10^{13}$ M$_\odot$. 
The N and NW peaks are recovered at positions consistent with \citetalias{Jauzac2016}. Although \citetalias{Jauzac2016} treated NW and S3 as separate substructures, they correspond here to a single NW region in this work, exhibiting two mass peaks associated with the two brightest cluster galaxies. For direct comparison, these peaks are labelled NW1 and NW2 in Table\,\ref{tab:a2744_subs}. However, both N and NW (NW and S3 in \citetalias{Jauzac2016}), reveal to be more massive in our analysis compared to \citetalias{Jauzac2016} as shown in Table\,\ref{tab:a2744_subs}. 
For W, \citetalias{Jauzac2016} detected a substructure, W$_{bis}$, which they identified as a probable projected background structure after identifying galaxies in its vicinity as background sources. However, the substructure found in this work, is farther to the West from W$_{bis}$, and is associated with cluster galaxies. 
The S1 substructure, located North-East of the core, is detected at a position closely matching that of \citetalias{Jauzac2016} and aligns with the gas bridge reported by \citet{Eckert2015}. The additional components S2 (North of S1 and East of N) and S4 (West of NW), reported by \citetalias{Jauzac2016}, are not recovered in this analysis. 
These non-detections may be attributable to the cluster-scale halos in the N and NW regions in our model more effectively incorporate the different mass contributions in these areas which may explain the non-detections of S2 and S4. The reconstructed mass distribution also shows elongations towards the southern and northwestern directions. These features are broadly consistent with the directions of the filamentary structures surrounding Abell,2744 reported in previous studies \citep{Eckert2015,Gallo2024,Cha2026}.
As discussed in Sect.\,\ref{sec:a2744_comparing_previous_work}, the strong+weak lensing model in this work yields more mass in the outskirts compared to previous studies. This effect is also observed in substructures in the outskirts. It is noted that substructures in Table\,\ref{tab:a2744_subs} corresponding to this work are more massive than in \citetalias{Jauzac2016}. Although, the mass of the core is within error bars of \citetalias{Jauzac2016}. 

\section{BAHAMAS ANALYSIS}
\label{sec:bahamas_analysis}

To provide a consistent two-dimensional comparison between Abell\,2744 and the BAHAMAS simulations, we analyse projected mass maps using the same wavelet based substructure detection procedure and selection criteria. Rather than using the native three-dimensional subhalo catalogues from the simulations, we identify substructures directly in the projected maps. We note, however, that the underlying mass maps are constructed differently. The Abell\,2744 mass map is reconstructed from strong and weak-lensing constraints using a multi-scale RBF grid informed by the observed light distribution, whereas the simulated maps are constructed directly from projected particle masses. This difference represents a limitation of the comparison and is taken into account when interpreting the resulting subhalo mass functions.

For each dark matter model, we analyse ten simulated galaxy clusters. Each cluster is projected along three orthogonal axes, producing three independent projected mass maps per halo. This results in an effective sample of thirty line-of-sight (LOS) clusters per dark matter model. All statistical quantities for the simulations are therefore computed using these thirty projected realisations, providing a robust ensemble for estimating mean trends and associated uncertainties.

For each LOS, a projected total mass map is constructed by depositing particle masses onto a two-dimensional grid using a cloud-in-cell assignment scheme. The field of view is fixed in physical units to match that used for Abell\,2744, and the angular pixel scale is chosen to be identical to the lensing map sampling. The resulting surface density maps are converted into the same internal units used for the Abell\,2744 wavelet analysis, allowing the same detection code to be applied without modification. To ensure that the observational and simulated analyses probe the same effective area, we apply a mask corresponding to the BUFFALO main-field footprint. 

The wavelet transform detection algorithm described in Sect.~\ref{sec:substructure_detection} is then applied to each simulated map using identical parameter choices. In particular, the wavelet scale is matched in physical units between Abell\,2744 and the simulations, ensuring sensitivity to substructures at the same characteristic spatial scale. Candidate peaks are identified from the thresholded wavelet coefficient map and iteratively refined by recentering circular apertures on the local maximum coefficient. A minimum projected separation criterion is imposed so that only the most massive detection is retained within a specified radius, preventing multiple detections of the same overdensity.

For each retained peak, the substructure mass is defined consistently as the projected mass enclosed within a circular aperture of 150\,kpc. In addition, for each LOS, we compute a projected host mass within a fixed 900\,kpc aperture centered on the cluster. This host mass is used to construct a dimensionless substructure to host mass ratio, enabling a second representation of the substructure population that partially accounts for cluster to cluster variations in overall mass normalisation. These peak catalogues are used to construct a subhalo mass function which is discussed in detail in Section\,\ref{sec:shmf}.

\section{DISCUSSION}
\label{sec:discussion}

\subsection{Comparison between different lensing mass models}
\label{sec:a2744_comparing_this_work}
Three main mass models are presented in this analysis: a parametric model for the cluster core, a non-parametric grid based model for the outskirts, and a combined model that integrates both strong and weak-lensing constraints across the entire field. Resulting mass distributions from these models are shown in Fig.\,\ref{fig:a2744_contours}. The inner core and prominent substructures to the North and North-West (N and NW) are well traced by the parametric model and are also visible, with less defined contours in the grid-only model.

Figure~\ref{fig:a2744_sdp1} presents the two-dimensional surface density profiles derived from each model. As observed, the amplitude of the grid-only model remains consistently lower than that of the parametric and combined models across all scales in the cluster. This discrepancy arises due to the absence of strong-lensing constraints in the grid-only model and the lack of an explicit large scale cluster halo component.

In contrast, the combined model exhibits systematically higher amplitudes compared to individual models, particularly in the outer regions of the cluster. Notably, the total mass inferred from the combined model exceeds the sum of the parametric and grid-only contributions in the outskirts. Ideally, in a sequential fit approach, the mass distribution derived from the parametric model should be accounted for during the grid optimisation step, allowing for the grid amplitudes to be adjusted accordingly. However, in practice, the combined model appears to overestimate the mass in the outskirts. This effect warrants further investigation, using simulations to better understand the origin of the discrepancy. A similar issue was reported by \citet{Niemiec2023}, who also found an overestimation of mass in the outer regions for the combined lensing models from both joint and sequential fit in \lenstool. This overestimation does not affect the positions of the detected substructures or the composition of the sample, but the measured aperture masses of the outer substructures may be biased high as a result, since the bias grows towards the outskirts while the central mass is essentially unaffected (see Sect.\,\ref{sec:shmf}).

\begin{table*}
\caption{A summary table to compare the mass of Abell\,2744 with previous works. Masses measured within 200\,kpc, 250\,kpc, 1000\,kpc and 1300\,kpc using different methods are quoted in units of $10^{14}$\,M$_{\odot}$. Other works are listed : \citet[][J15]{Jauzac2015_a2744} and \citet[][J16]{Jauzac2016}, \citet[][F23]{Furtak2023b}, and \citet[][C24]{Cha2024}.}
\centering
\makebox[\textwidth][c]{
\begin{tabular}{c c c c c c c}
\hline\hline
 & Method & J15 & J16 & F23 & C24 & This work \\
\hline\hline
\textit{R} < 200\,kpc & SL only & 2.16 $\pm$ 0.05 & - & 1.60 $\pm$ 0.22 & - & 1.69 $\pm$ 0.03\\
             & SL + WL & - & - & - & 1.73 & 2.01 $\pm$ 0.05 \\
\hline
\textit{R} < 250\,kpc & SL only & 2.76 $\pm$ 0.08 & - & - & - & 2.22 $\pm$ 0.04  \\
             & SL + WL & - & 2.77 $\pm$ 0.01 & - & - &  2.73 $\pm$ 0.09 \\
\hline
\textit{R} < 1000\,kpc & WL only & - & - & - & - & 13.24 $\pm$ 1.07 \\
             & SL + WL & - & 18.5 $\pm$ 0.7 & - & 11.9 & 25.0 $\pm$ 2.0 \\
\hline
\textit{R} < 1300\,kpc & WL only & - & - & - & - & 16.71 $\pm$ 1.57 \\
             & SL + WL & - & 23$\pm$1 & - & - & 33.86 $\pm$ 3.4 \\
\hline\hline
\end{tabular}}
\label{tab:a2744_masses}
\end{table*}

\subsection{Comparison with Previous Work}
\label{sec:a2744_comparing_previous_work}

\citet{Jauzac2015_a2744} performed a strong-lensing analysis of Abell\,2744 with deep HFF observations using \lenstool. They found a projected mass within 250\,kpc of $M_{\mathrm{J15}}(R < 250\,\mathrm{kpc}) = (2.76 \pm 0.08) \times 10^{14}$ M$_\odot$, which is slightly larger than the strong-lensing mass in this work, $M_{\mathrm{SL}}(R < 250\,\mathrm{kpc}) = (2.22 \pm 0.04) \times 10^{14}$ M$_\odot$.

From the strong+weak-lensing analysis with HFF in the core with \lenstool\ by \citetalias{Jauzac2016}, they found a projected mass within 250\,kpc of $M_{\mathrm{J16}}(R < 250\,\mathrm{kpc}) = (2.77 \pm 0.01) \times 10^{14}$ M$_\odot$, which is within the errorbars of the strong+weak-lensing mass obtained in this work, $M_{\mathrm{SL+WL}}(R < 250\,\mathrm{kpc}) = (2.73 \pm 0.06) \times 10^{14}$ M$_\odot$. \citetalias{Jauzac2016} presented a projected mass within 1.3\,Mpc of $M_{\mathrm{J16}}(R < 1.3\,\mathrm{Mpc}) = (2.3 \pm 0.1) \times 10^{15}$ M$_\odot$, while this work gets a much higher mass, $M_{\mathrm{SL+WL}}(R < 1.3\,\mathrm{Mpc}) = (3.38 \pm 0.34) \times 10^{15}$ M$_\odot$. The mass from this work is much higher than that of \citetalias{Jauzac2016}, due to overestimation of our strong+weak lensing mass, as mentioned in the previous section.

From the \textit{JWST}/UNCOVER observations, \citet{Furtak2023a} performed a strong-lensing analysis of Abell\,2744, yielding a projected mass of $M_{\mathrm{F23}}(R < 200\,\mathrm{kpc}) = (1.60 \pm 0.22) \times 10^{14}$ M$_\odot$, which is within errorbars of strong-lensing model in this work, $M_{\mathrm{SL}}(R < 200\,\mathrm{kpc}) = (1.69 \pm 0.03) \times 10^{14}$ M$_\odot$, but lower than the combined model but still within its errorbars, $M_{\mathrm{SL+WL}}(R < 200\,\mathrm{kpc}) = (2.01 \pm 0.06) \times 10^{14}$ M$_\odot$. 

One of the very first weak-lensing only analyses with \jwst\ of Abell\,2744 was done by \citet{Harvey2024} using a non-parametric method, with a background source density of $\sim$150 galaxies.arcmin$^{-2}$ compared to $\sim$60 galaxies.arcmin$^{-2}$ from \textit{JWST} in this work. The difference in the source density comes from the method used to obtain the weak-lensing catalogue. In this analysis background sources with spectroscopic redshifts and secured photometric redshifts are taken into account, while \citet{Harvey2024} makes the lensing cuts mentioned in \ref{sec:jwst_wlcat}, but did not perform a colour-colour selection.
\citet{Harvey2024} finds a mass of $M_{\mathrm{H24}}(R < 250\,\mathrm{kpc}) = (1.60 \pm 0.06) \times 10^{14}$ M$_\odot$, which is within error bars of the weak-lensing mass we obtain in this work $M_{\mathrm{WL}}(R < 250\,\mathrm{kpc}) = (1.77 \pm 0.09) \times 10^{14}$ M$_\odot$. It should be noted that masses derived from weak‐lensing alone and confined to the core region (250\,kpc) are systematically lower than those from strong‐lensing or combined models, due to the lack of lensing constraints in the central cluster core. 

Another recent \jwst\ strong+weak-lensing analysis was performed by \cite{Cha2024}. In that study, a free‐form maximum entropy reconstruction (MARS) algorithm was used, in contrast to the combination of parametric and non‐parametric methods used here. A weak‐lensing source density of $\sim$350 galaxies.\,arcmin$^{-2}$ was obtained, owing to a background selection based solely on photometric redshifts without further cuts. \citet{Cha2024} reported a projected mass of $M_{\mathrm{SL+WL}}(R < 1000\,\mathrm{kpc}) = (1.19) \times 10^{15}$ M$_\odot$, which is much lower compared to this work, $M_{\mathrm{SL+WL}}(R < 1000\,\mathrm{kpc}) = (2.50 \pm 0.20) \times 10^{15}$ M$_\odot$. However, it is comparable to the strong-lensing only and weak-lensing only mass from this work, $M_{\mathrm{SL}}(R < 1000\,\mathrm{kpc}) = (1.12 \pm 0.03) \times 10^{15}$ M$_\odot$ and $M_{\mathrm{WL}}(R < 1000\,\mathrm{kpc}) = (1.32 \pm 0.11) \times 10^{15}$ M$_\odot$ respectively.

Although both \citet{Harvey2024} and \citet{Cha2024} recover the three main structures in the cluster, core, N and NW, but they have no evidence of other substructures detected in the field, in contrast to those identified and reported by \citetalias{Jauzac2016} and in this work. A summary of the different mass estimates is presented in Table\,\ref{tab:a2744_masses}.

\subsection{Subhalo mass function (SHMF)}
\label{sec:shmf}
The range of substructure masses and their spatial distribution in dynamically complex, non-relaxed clusters such as Abell\,2744 provides a valuable opportunity to probe the subhalo mass function (SHMF) across diverse environments. The SHMF characterises the number of dark matter subhalos embedded within the host halo of a galaxy cluster as a function of their mass. It serves not only as a critical diagnostic of hierarchical structure formation but also as a direct observational probe of the underlying nature of dark matter.

Within the standard $\Lambda$CDM framework, the SHMF is predicted to follow a steep power law distribution, indicating a high abundance of low mass subhalos. However, various theoretical uncertainties and observational challenges such as the missing satellites problem and the cusp-core discrepancy, suggest that the $\Lambda$CDM prediction may be incomplete. To address these issues, alternative dark matter models such as WDM and SIDM have been proposed, each offering distinct predictions for the SHMF. For instance, in WDM scenarios, a sharp suppression of low mass subhalos is expected below approximately $10^8$–$10^9$\,M$_\odot$ \citep{lovell2014,Kennedy2014}, while SIDM models predict a 30–50$\%$ reduction in the abundance of centrally concentrated low mass subhalos, potentially mitigating observed tensions more effectively \citep{Vogelsberger2012,Rocha2013}.

Previous studies have shown that both WDM and SIDM models produce fewer low-mass subhalos compared to CDM scenarios \citep{Springel2008, lovell2014}. However, these distinctions are primarily explored in the low-mass regime of the SHMF, where differences between dark matter models are expected to be most pronounced. In this work, we instead focus on the higher-mass end of the SHMF, which is directly accessible with current strong and weak-lensing observations of massive clusters such as Abell\,2744.

Here, we explore the high-mass regime of the SHMF and investigate how it is impacted by different dark matter models and directly compare these theoretical SHMF with our observational results. Throughout this work, the SHMFs refer to projected (2D) subhalo mass functions rather than the intrinsic three-dimensional distributions.
We construct a SHMF comparison between observations and simulations using the substructures detected with the WT method as discussed in Section\,\ref{sec:substructure_detection} and \ref{sec:bahamas_analysis}.

\begin{figure*}
  \centering
  \subfloat[]{
    \includegraphics[width=0.48\linewidth]{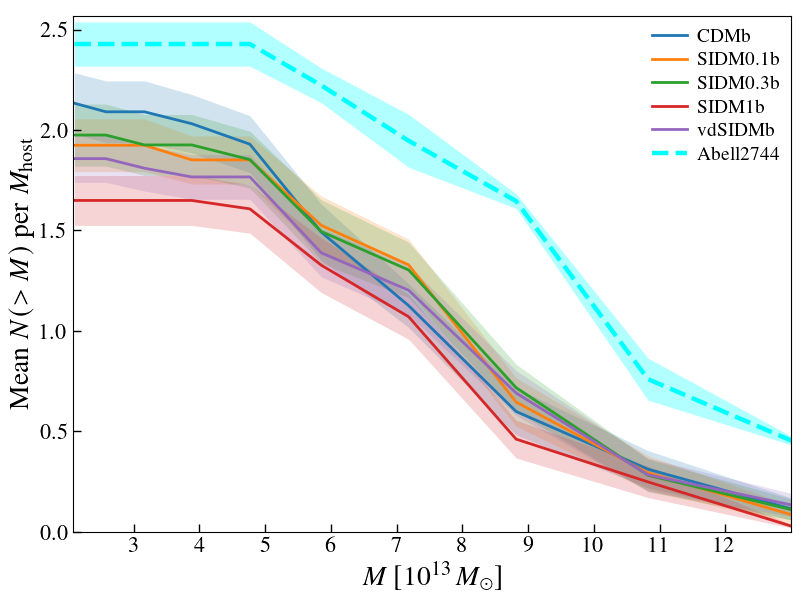}
    \label{fig:shmf1}
  }
  \hfill
  \subfloat[]{
    \includegraphics[width=0.48\linewidth]{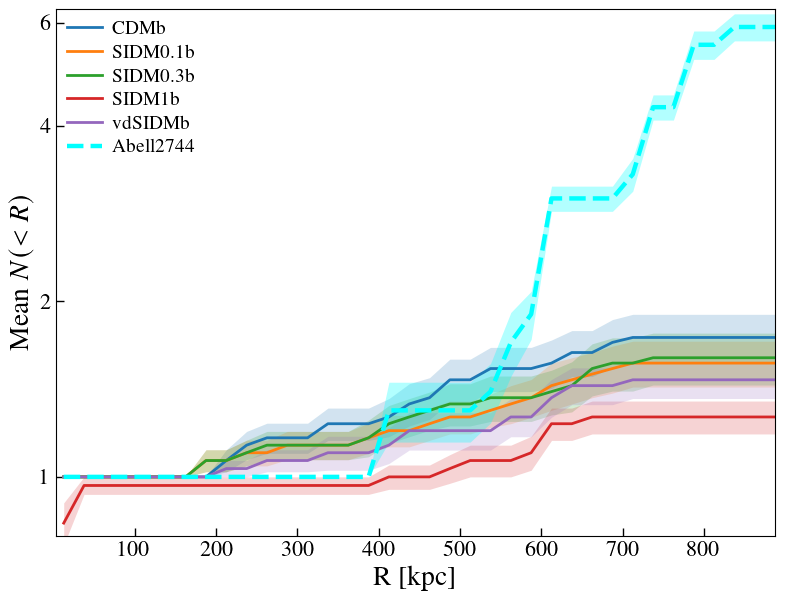}
    \label{fig:shmf2}
  }
  \caption{ 
  \textit{Left:} Host mass normalised cumulative subhalo mass function for Abell\,2744 and the BAHAMAS dark matter models (CDM, SIDM with cross section of 0.1, 0.3, and 1 cm$^2$g$^{-1}$, and velocity dependent SIDM). The curves show the mean number of detected substructures above a given aperture mass, normalised by the projected host mass within 900\,kpc and scaled per $10^{15}\,\mathrm{M_\odot}$. 
  \textit{Right:} Mean number of detected substructures as a function of projected radius from the cluster centre for Abell\,2744 (cyan) and five different dark matter models from BAHAMAS. }
  \label{fig:shmf}
\end{figure*}

The two diagnostics used to compare Abell\,2744 with the BAHAMAS simulations are the SHMF normalised by the cluster mass (left panel in Fig.\,\ref{fig:shmf}) and the cumulative radial distribution of detected substructures (right panel in Fig.\, \ref{fig:shmf}). The host mass normalised SHMF quantifies the mean number of substructures above a given aperture mass per unit host mass within 900\,kpc, thereby removing first order differences in overall cluster mass normalisation and isolating differences in substructure abundance and survival. All models exhibit the expected declining behaviour with increasing substructure mass, reflecting the rarity of high mass substructures. At the low mass end, the simulated CDM and SIDM models largely overlap, while differences become more apparent toward intermediate and high masses. The collisionless CDM model consistently produces the highest abundance of massive substructures among the simulations, whereas the strongest self-interacting model (SIDM1b) shows the most pronounced suppression. The velocity dependent SIDM case lies between these extremes, consistent with reduced interaction strength at cluster scale velocities. This ordering is physically expected: increasing the self-interaction cross section enhances core formation and tidal stripping, thereby reducing the number of substructures that remain sufficiently massive to be detected at the adopted wavelet scale and aperture definition.

In contrast, Abell\,2744 lies systematically above all simulated models across most of the mass range, even after normalisation by host mass. This demonstrates that the excess of substructures in Abell\,2744 is not driven simply by its overall mass, but reflects a higher abundance of detectable substructures per unit host mass. The discrepancy persists across the intermediate and high mass regime, where the simulations diverge most strongly between CDM and SIDM. Notably, increasing the self-interaction strength does not move the simulations toward the observational result. Instead, it further suppresses the predicted abundance.

The cumulative radial distribution provides complementary information on the spatial distribution of substructures. All simulated models show a gradual increase in the mean number of detected substructures with projected radius, reflecting the declining efficiency of tidal disruption toward the outskirts. The relative ordering of the models mirrors that seen in the SHMF: CDM yields the largest cumulative counts at most radii, while the strongest SIDM model yields the smallest. Differences between models become more pronounced beyond several hundred kpcs, where the cumulative counts diverge steadily. Abell\,2744 exhibits a markedly steeper rise in cumulative counts beyond approximately 600\,kpc compared to all simulations, reaching significantly higher total counts by 1\,Mpc. This indicates that the excess substructure abundance in Abell\,2744 is particularly prominent at intermediate to large radii.

Taken together, these two diagnostics show that Abell\,2744 contains both a higher abundance of massive substructures per unit host mass and a steeper radial accumulation of substructures than predicted by any of the simulated dark matter models analysed here, including CDM. The fact that the discrepancy remains after host mass normalisation suggests that it cannot be attributed solely to differences in overall cluster mass. Instead, it likely reflects the dynamically disturbed state of Abell\,2744, which is known to be undergoing complex merger activity. In such a configuration, infalling groups and smaller clusters may remain spatially distinct and detectable in projection before being fully disrupted. Because the same two-dimensional detection procedure, spatial scale, and aperture mass definition are applied to both the simulations and the observations, the differences highlighted in these plots are not attributable to the substructure detection method. We note, however, that the underlying mass maps are constructed differently (Sect.\,\ref{sec:bahamas_analysis}), so the comparison is not free of all methodological differences.

We note that the overestimation of mass in the cluster outskirts may bias the aperture masses of the outer substructures high. This does not change the detected sample, since all detections lie well above the selection threshold, and its effect on the normalised subhalo mass function is partly compensated, because the host mass used in the normalisation is measured from the same map. At the low-mass end, where Abell\,2744 lies only modestly above the CDM prediction, the excess is more sensitive to this systematic. However, it is important to emphasise that Abell\,2744 is an exceptional and dynamically extreme system. It is one of the most massive and morphologically disturbed clusters known, characterised by multiple ongoing mergers and infalling subclusters distributed over megaparsec scales. The elevated abundance of detected substructures per unit host mass may therefore reflect, in significant part, its disturbed dynamical state, rather than a fundamental failure of the underlying cosmological model. This apparent excess is also likely influenced by the fact that the comparison clusters were selected purely by mass, and not for a merger state comparable to that of Abell\,2744. The clusters used in this comparison are the most massive ones from the BAHAMAS suite, but they are not necessarily undergoing merger activity of comparable complexity or observed at a similar dynamical phase. 
We caution that this comparison, based on the most massive BAHAMAS clusters alone, is not sufficient to establish a definitive connection between the observed excess and the merger state of Abell\,2744. A larger simulated sample spanning a range of dynamical states would be required for a robust interpretation. Indeed, \citet{Wang2026} identified several Abell 2744 analogues in a much larger CDM simulation, showing that similarly complex multi-clump systems can be reproduced within $\Lambda$CDM given a sufficiently large volume.

Even with a consistent comparison method, some differences are therefore expected due to the natural variation in how galaxy clusters grow and evolve. In addition, as discussed in Sect\,\ref{sec:wl}, \ref{sec:sl+wl} and \ref{sec:bahamas_analysis}, the observed and simulated mass maps are constructed differently: the Abell\,2744 map is a lensing reconstruction on a light-informed multi-scale grid, whereas the simulated maps are direct projections of particle mass. This difference is a further reason to interpret the comparison with caution rather than as evidence for a discrepancy. In this context, the fact that Abell\,2744 lies above the mean simulated relations should not be interpreted as evidence for a discrepancy between theory and observations. Rather, it highlights that dynamically extreme clusters populate the upper tail of the substructure distribution predicted within $\Lambda$CDM. 
In parallel, extending this analysis to a larger sample of observed clusters spanning a range of dynamical states would be highly valuable. Applying the same methodology to both relaxed and disturbed systems would enable a more statistically robust comparison between observations and simulations, and help quantify how frequently clusters with Abell\,2744 like substructure richness arise within the $\Lambda$CDM framework.

\section{SUMMARY $\&$ CONCLUSION}
\label{sec:summary}
This paper presents a combined strong and weak-lensing mass model of the massive non-relaxed cluster, Abell\,2744 at $z = 0.308$, one of the six HFF and BUFFALO clusters. Our final model is made using the sequential fit by combining a parametric model using strong-lensing constraints in the core and a non-parametric grid model using weak-lensing constraints in the outskirts. 

The strong-lensing model is optimised using 37 spectroscopically confirmed multiple image systems in the core, N, and NW regions from \hst/HFF and BUFFALO combined with \jwst/UNCOVER observations. All components are optimized simultaneously to obtain a precise mass distribution not only of the core but also of the N and NW regions, which were poorly constrained prior to \jwst~observations due to lack of identifications of multiple images in the \textit{HST} observations. Our best-fit mass model achieves an rms of 0.83\arcsec\ with a projected enclosed mass of $M_{\mathrm{SL}}(R < 200,\mathrm{kpc}) = (1.69 \pm 0.03) \times 10^{14}$ M$_\odot$ in the core.

In the outskirts, the model is optimised using weakly lensed background galaxies as constraints within a non-parametric grid framework. Combining constraints from \textit{HST}/BUFFALO and \textit{JWST}/UNCOVER observations provides a total of 5209 background galaxies, yielding a source density of $\sim$66 galaxies per arcmin$^2$. This is the first time that a combined \hst\ and \jwst\ weak-lensing analysis is performed. This weak-lensing only model yields a projected mass of $M_{\mathrm{WL}}(R < 1000\,\mathrm{kpc}) = (1.32 \pm 0.11) \times 10^{15}$ M$_\odot$.

The final model consists of a sequential fit which combines both strong and weak lensing constraints. First, the parametric model is fixed to the best-fit parameters, and a non-parametric grid is added to model the outskirts of the cluster. The combined model yields a mass of $M_{\mathrm{SL+WL}}(R < 200\,\mathrm{kpc}) = (2.01 \pm 0.06) \times 10^{14}$ M$_\odot$ in the core, and an enclosed mass of $M_{\mathrm{SL+WL}}(R < 1000\,\mathrm{kpc}) = (2.50 \pm 0.20) \times 10^{15}$ M$_\odot$ within 1000\,kpc from the center of the cluster. 

Massive galaxy clusters with substantial substructure, such as Abell\,2744, are recognised as valuable laboratories for testing the $\Lambda$CDM model of cosmology \citep{Jauzac2016,Schwinn2017,schwinn2018,Jauzac2018}. Through detailed gravitational lensing analyses, evidence for numerous massive substructures can be revealed. These features trace the hierarchical assembly of clusters and offer observational probes of the dark matter distribution.

Using a wavelet-based detection technique initially presented in \citep{schwinn2018}, we constructed a host-mass normalised subhalo mass function (SHMF) and cumulative radial distribution for Abell\,2744. We then performed comparison with massive clusters from the BAHAMAS hydrodynamical simulations by applying the same wavelet-based detection method to both observational and simulated projected mass maps. We note, however, that the observational mass map is reconstructed using a multi-scale RBF grid informed by the cluster light distribution, whereas the simulated maps are constructed directly from projected particle masses. This difference in mass map reconstruction means the comparison, while methodologically consistent at the detection stage, is not fully independent of the light distribution used to build the observational grid. This represents one of the first SHMF comparisons between lensing-derived mass maps and cosmological hydrodynamical simulations using a broadly consistent 2D substructure-detection framework.

The comparison shows that Abell\,2744 exhibits a higher abundance of detectable substructures per unit host mass than the mean prediction of the simulated CDM and SIDM models considered. However, this result must be interpreted with caution. Abell\,2744 is a dynamically extreme, highly disturbed merging system and does not represent a typical relaxed cluster. One of the possible reasons for its elevated substructure content is its transient merger phase, rather than a fundamental discrepancy with $\Lambda$CDM. However, we note that the BAHAMAS clusters used for comparison, while the most massive available in the simulation suite, are not necessarily undergoing merger activity of comparable complexity to Abell\,2744. Simulations such as those of \citet{Wang2026}, which identify Abell\,2744 like analogues with complex multi-clump configurations within a $\Lambda$CDM framework, suggest that similarly disturbed systems can arise in simulations, and a more targeted comparison with such analogues would help to better isolate the role of dynamical state in driving the observed excess. In hierarchical structure formation, such systems are expected to populate the upper tail of the substructure distribution. Therefore, the results presented here do not constitute evidence for tension between observations and theory, but rather demonstrate the importance of accounting for dynamical state when interpreting SHMF measurements together with the need for simulated highly-disturbed clusters.

The strengths of the BAHAMAS comparison lie in the controlled cosmological framework and the availability of both CDM and SIDM variants, allowing for a physically motivated exploration of how self-interactions modify subhalo survival and radial distributions. The host-mass normalisation further removes first-order differences in overall cluster mass, isolating variations in substructure abundance. However, the simulation sample remains limited in the number of extremely massive, dynamically complex clusters comparable to Abell\,2744. While BAHAMAS provides statistically meaningful projections through multiple lines of sight, a larger ensemble of very massive and highly disturbed clusters  would be required to robustly characterise the high-mass tail of the SHMF distribution.

On the observational side, the primary limitation is that this analysis is based on a single cluster. Although Abell\,2744 is uniquely well constrained due to deep \hst~and \jwst~data, a statistically robust assessment of the SHMF requires a larger sample of clusters spanning a range of masses and dynamical states. Future wide-field surveys such as \textit{Euclid} and dedicated strong-lensing surveys such as Strong LensIng and Cluster Evolution \citep[SLICE, GO:\#5594, PI: G. Mahler;][]{Cerny2025,Allingham2025,Limousin2026} will dramatically increase the number of massive clusters with high-quality weak and strong-lensing constraints. These datasets will enable systematic SHMF measurements across a statistically significant sample, allowing the intrinsic scatter due to merger activity and assembly history to be quantified.

On the simulation side, next-generation large-volume simulations containing a larger number of $M \gtrsim 10^{15}\,\mathrm{M_\odot}$ clusters are becoming essential. While simulations such as TNG-Clusters provide a substantial population of massive systems, they are currently limited to CDM-only cosmologies. Extending such large volume cluster simulations to include alternative dark matter models would enable more comprehensive SHMF comparisons across a broader parameter space.

In summary, this work demonstrates the power of combining deep \hst~and \jwst~lensing constraints to construct high-resolution mass maps and perform consistent SHMF measurements. Abell\,2744 provides a proof of concept for such analyses, but it remains an exceptional and dynamically complex system. A statistically meaningful confrontation between observations and dark matter models now requires both a larger observational sample and expanded simulation suites containing numerous massive clusters across different dynamical states and incorporating different dark matter candidates.

\section*{Acknowledgements}
NP is supported by the Science and Technology Facilities Council (STFC, grant numbers ST/S505365/1, ST/V506643/1 and ST/W507428/1). NP, MJ and DL acknowledge support from the United Kingdom Research and Innovation (UKRI) Future Leaders Fellowship `Using Cosmic Beasts to uncover the Nature of Dark Matter' (grant number MR/X006069/1). SB is supported by the UK Research and Innovation (UKRI) Future Leaders Fellowship(Grant No. MR/V023381/1 and UKRI2044). ML acknowledges the Centre National de la Recherche Scientifique (CNRS) and the Centre National des Etudes Spatiale (CNES) for support.
The authors thank Matthias Bartelmann for his assistance in recovering the original wavelet transform code used in \citet{schwinn2018}, which formed the basis of the adapted implementation used in this work.

\section*{Data Availability}
The BUFFALO \textit{HST} data used for this work are available on the MAST archive in the BUFFALO repository for Abell\,2744 \footnote{https://archive.stsci.edu/pub/hlsp/buffalo/internal/data/abell2744/}. The UNCOVER \textit{JWST} data used in this work are available on the UNCOVER website \footnote{https://jwst-uncover.github.io/}. The lensing mass models and other data products are available at the Strong Lensing Cluster Atlas Data Base, which is hosted at Laboratoire
d’Astrophysique de Marseille \footnote{https://data.lam.fr/sl-cluster-atlas/home}. 

\bibliographystyle{mnras}
\bibliography{ref.bib} 


\appendix
\setcounter{table}{0}
\renewcommand{\thetable}{A\arabic{table}}
\begin{table*}
\centering
\caption{List of observations of Abell\,2744 used in this work. Columns 1 and 2 give the right ascension and declination of the observed field in degrees. Column 3 lists the instruments and filters. Column 4 and 5 give the exposure time in seconds and the observation date, respectively. Finally, columns 6 and 7 provide the programme ID and the PI.}
\makebox[\textwidth][c]{
\label{tab:obs}
\begin{tabular}{c c c c c c c c}
\hline\hline
Programme/Survey &  Instrument  & Date & Exp. Time (s) &  Prog. ID & PI \\
\hline\hline
\multicolumn{6}{c}{\textbf{\textit{HST}}} \\
\hline
HFF &  WFC3/F105W & 25/10/13-01/07/14 & 24.5 orbits$^\ast$ & 13495 & Lotz  \\
    &  WFC3/F125W & &  12 orbits & & \\
    &  WFC3/F140W & &  10 orbits &  &\\
    &  WFC3/F160W & &  14.5 orbits & & \\
    &  ACS/F435W & & 24 orbits &  &  \\
    &  ACS/F606W & &  14 orbits &  &\\
    &  ACS/F814W & &  46 orbits &  &\\
\hline
BUFFALO & ACS/F606W & 15/05/19-07/12/19 & 1538 & 15117 & Steinhardt\\
        & ACS/F814W & & 2934 &   & \\
        & WFC3/F105W & & 1426 &  & \\
        & WFC3/F125W & & 1467 &  & \\
        & WFC3/F160W & & 1579 &  & \\
\hline\hline
\multicolumn{6}{c}{\textbf{\textit{JWST}}} \\
\hline
UNCOVER & NIRCam/F115W & 2/11/22-15/11/22 & 21600 & 2561 & Labbe \\
 & NIRCam/F115W &  & 21600 &  &  \\
 & NIRCam/F150W &  &13320&  &  \\
 & NIRCam/F200W &  &13320&  &  \\
 & NIRCam/F277W &  &13320&  &  \\
 & NIRCam/F356W &  &13320&  &  \\
 & NIRCam/F410W &  &13320&  &  \\
 & NIRCam/F444W &  &16560&  &  \\
 & NIRISS/F115W &  &13320&  &  \\
 & NIRISS/F150W &  &13320&  &  \\
 & NIRISS/F200W &  &13320&  &  \\
 & NIRISS/F356W &  &7560&  &  \\
 & NIRISS/F444W &  &7560&  &  \\
 & NIRCam/F090W & 01/07/23-30/08/23 &19080&  &  \\
 & NIRCam/F115W &  &19080&  &  \\
 & NIRCam/F150W &  &19080&  &  \\
 & NIRCam/F200W &  &9360&  &  \\
 & NIRCam/F277W &  &9360&  &  \\
 & NIRCam/F335W &  &9360&  &  \\
 & NIRCam/F356W &  &9360&  &  \\
 & NIRCam/F410W &  &9360&  &  \\
 & NIRCam/F444W &  &19080&  &  \\
\hline\hline
\multicolumn{6}{l}{\footnotesize{$^\ast$ Due to the large exposure time of HFF, exposure times are reported in orbits rather than seconds.}}\\
\end{tabular}}
\end{table*}

\begin{table}
\centering
\small
\caption{Multiple-image systems used as constraints in the strong-lensing analysis of Abell\,2744.}
\label{tab:a2744_arcs}
\begin{tabular}{lccc}
\hline
ID & R.A. (deg) & Dec. (deg) & $z_{\mathrm{spec}}$ \\
\hline
3.1 & 3.589625 & -30.393906 & 3.9803 \\
3.2 & 3.588429 & -30.393786 & 3.9803 \\
4.1 & 3.592115 & -30.402634 & 3.5719 \\
4.2 & 3.595643 & -30.401623 & 3.5719 \\
4.4 & 3.593193 & -30.404915 & 3.5719 \\
4.5 & 3.593593 & -30.405106 & 3.5719 \\
5.2 & 3.584982 & -30.391374 & 4.0225 \\
5.3 & 3.579958 & -30.394772 & 4.0225 \\
105.1 & 3.583430 & -30.392070 & 4.0225 \\
105.2 & 3.582292 & -30.392789 & 4.0225 \\
8.1 & 3.589698 & -30.394339 & 3.9750 \\
8.2 & 3.588815 & -30.394223 & 3.9750 \\
10.1 & 3.588417 & -30.405878 & 2.6550 \\
10.2 & 3.587375 & -30.406481 & 2.6550 \\
18.1 & 3.576123 & -30.404488 & 5.6605 \\
18.2 & 3.588379 & -30.395649 & 5.6605 \\
18.3 & 3.590711 & -30.395561 & 5.6605 \\
22.1 & 3.587917 & -30.411611 & 5.2830 \\
22.2 & 3.600083 & -30.404417 & 5.2830 \\
22.3 & 3.596542 & -30.409031 & 5.2830 \\
24.1 & 3.595900 & -30.404480 & 1.0430 \\
24.2 & 3.595125 & -30.405933 & 1.0430 \\
24.3 & 3.587333 & -30.409103 & 1.0430 \\
26.1 & 3.593898 & -30.409731 & 3.0537 \\
26.2 & 3.590346 & -30.410581 & 3.0537 \\
26.3 & 3.600110 & -30.402942 & 3.0537 \\
30.1 & 3.591000 & -30.397444 & 1.0250 \\
30.2 & 3.586708 & -30.398186 & 1.0250 \\
31.1 & 3.585917 & -30.403167 & 4.7584 \\
31.2 & 3.583708 & -30.404117 & 4.7584 \\
33.1 & 3.584708 & -30.403147 & 5.7235 \\
33.2 & 3.584396 & -30.403400 & 5.7235 \\
34.1 & 3.593412 & -30.410842 & 3.7840 \\
34.2 & 3.593817 & -30.410725 & 3.7840 \\
34.3 & 3.600705 & -30.404605 & 3.7840 \\
37.1 & 3.589042 & -30.394914 & 2.6490 \\
37.2 & 3.588708 & -30.394853 & 2.6490 \\
39.1 & 3.588792 & -30.392530 & 4.0150 \\
39.2 & 3.588542 & -30.392508 & 4.0150 \\
40.1 & 3.589083 & -30.392664 & 4.0000 \\
40.2 & 3.588208 & -30.392553 & 4.0000 \\
41.1 & 3.599176 & -30.399582 & 4.9113 \\
41.2 & 3.593557 & -30.407769 & 4.9113 \\
41.3 & 3.583447 & -30.408500 & 4.9113 \\
41.4 & 3.590617 & -30.404459 & 4.9113 \\
42.1 & 3.597313 & -30.400608 & 3.6900 \\
42.2 & 3.590958 & -30.403261 & 3.6900 \\
42.3 & 3.581583 & -30.408633 & 3.6900 \\
42.4 & 3.594250 & -30.406389 & 3.6900 \\
42.5 & 3.592413 & -30.405194 & 3.6900 \\
47.1 & 3.590162 & -30.392181 & 4.0225 \\
47.2 & 3.585842 & -30.392244 & 4.0225 \\
47.3 & 3.578329 & -30.398133 & 4.0225 \\
147.1 & 3.589679 & -30.392136 & 4.0225 \\
147.2 & 3.586454 & -30.392128 & 4.0225 \\
147.3 & 3.578008 & -30.398392 & 4.0225 \\
\hline
\end{tabular}
\end{table}

\begin{table}
\centering
\small
\contcaption{Multiple-image systems used as constraints in the strong-lensing analysis of Abell\,2744.}
\begin{tabular}{lccc}
\hline
ID & R.A. (deg) & Dec. (deg) & $z_{\mathrm{spec}}$ \\
\hline
61.1 & 3.595533 & -30.403499 & 2.9520 \\
61.2 & 3.595143 & -30.404495 & 2.9520 \\
62.1 & 3.591326 & -30.398643 & 4.1925 \\
62.2 & 3.590582 & -30.398918 & 4.1925 \\
63.1 & 3.582261 & -30.407166 & 5.6599 \\
63.2 & 3.592758 & -30.407022 & 5.6599 \\
63.3 & 3.589133 & -30.403419 & 5.6599 \\
63.4 & 3.598806 & -30.398279 & 5.6599 \\
64.1 & 3.581197 & -30.398708 & 3.4072 \\
64.3 & 3.596333 & -30.394232 & 3.4072 \\
55.1 & 3.597042 & -30.404746 & 1.6880 \\
55.2 & 3.596371 & -30.406161 & 1.6880 \\
55.3 & 3.585726 & -30.410085 & 1.6880 \\
A65.1 & 3.556540 & -30.375738 & 3.5194 \\
A65.2 & 3.556335 & -30.376786 & 3.5194 \\
B65.1 & 3.556544 & -30.375812 & 3.5194 \\
B65.2 & 3.556371 & -30.376715 & 3.5194 \\
C65.1 & 3.556508 & -30.376163 & 3.5194 \\
C65.2 & 3.556479 & -30.376392 & 3.5194 \\
67.1 & 3.542036 & -30.374609 & 2.3220 \\
67.2 & 3.542825 & -30.373502 & 2.3220 \\
67.3 & 3.544295 & -30.368063 & 2.3220 \\
68.1 & 3.540444 & -30.375069 & 2.5840 \\
68.2 & 3.540333 & -30.372494 & 2.5840 \\
68.3 & 3.543054 & -30.368098 & 2.5840 \\
68.4 & 3.542082 & -30.372626 & 2.5840 \\
69.1 & 3.540719 & -30.375880 & 2.4109 \\
69.2 & 3.540927 & -30.372222 & 2.4109 \\
69.3 & 3.543098 & -30.369107 & 2.4109 \\
69.4 & 3.542633 & -30.372462 & 2.4109 \\
70.1 & 3.552409 & -30.373992 & 2.3915 \\
70.2 & 3.552351 & -30.374079 & 2.3915 \\
72.1 & 3.549645 & -30.378464 & 3.7468 \\
72.2 & 3.551115 & -30.374758 & 3.7468 \\
72.3 & 3.552295 & -30.372243 & 3.7468 \\
74.1 & 3.545136 & -30.374918 & 2.3743 \\
74.2 & 3.545250 & -30.374700 & 2.3743 \\
74.3 & 3.547033 & -30.369489 & 2.3743 \\
A78.1 & 3.579614 & -30.357722 & 2.3151 \\
A78.2 & 3.579236 & -30.357810 & 2.3151 \\
A78.3 & 3.574876 & -30.356072 & 2.3151 \\
B78.1 & 3.579706 & -30.357722 & 2.3151 \\
B78.2 & 3.579162 & -30.357838 & 2.3151 \\
B78.3 & 3.574914 & -30.356079 & 2.3151 \\
\hline
\end{tabular}
\end{table}

\begin{table*}
\scriptsize
\caption{Best fit parameters for our strong lensing mass model of Abell\,2744. Values given in brackets are fixed in our model, while the ones quoted with error bars are being optimised. Coordinates of the cluster and BCG halos are given in arcseconds relative to the cluster center (R.A., Dec)$=$(3.586259 -30.400174). DM1 and DM2 corresponds to the cluster-scale dark matter haloes used in the core, DM3 and DM4, for NW halos and DM5 for N halo. The two BCGs in the core, optimised separately from the other cluster members are used as BCG1 and BCG2.} 
\centering
    \makebox[\textwidth][c]{
    \begin{tabular}{c c c c c c c c}
    \hline\hline 
    Components & $\Delta$R.A.(\arcsec) & $\Delta$Dec (\arcsec) & \textit{e} & $\theta$(deg) & $r_{\mathrm{core}}$ (kpc) & $r_{\mathrm{cut}}$ (kpc) & $\sigma_0$ (km\,s$^{-1}$)
    \\
    \hline
    DM1& 0.32 $\pm$ 0.13 & 2.13 $\pm$ 0.47 & 0.85 $\pm$ 0.02 & 92.65 $\pm$ 0.99 & 94.16 $\pm$ 3.77 & [1000] & 566.88 $\pm$ 12.27 \\
    \\
    DM2& -16.75 $\pm$ 0.09 & -15.02 $\pm$ 0.11 & 0.54 $\pm$ 0.01 & 50.62 $\pm$ 0.21 & 53.56 $\pm$ 2.83 & [1000] & 831.38 $\pm$ 4.36 \\
    \\
    DM3& 109.99 $\pm$ 1.65 & 90.35 $\pm$ 1.20 & 0.79 $\pm$ 0.03 & 30.22 $\pm$ 3.24 & 46.8 $\pm$ 24.96 & [1000] & 473.93 $\pm$ 77.27 \\
    \\
    DM4& 140.13 $\pm$ 1.94 & 101.16 $\pm$ 1.16 & 0.36 $\pm$ 0.12 & 8.13 $\pm$ 5.85 & 77.06 $\pm$ 21.62 & [1000] & 734.27 $\pm$ 99.34 \\
    \\
    DM5& [24.25] & [155.82] & 0.58 $\pm$ 0.04 & 31.82 $\pm$ 3.25 & 10.53 $\pm$ 2.89 & [1000] & 544.11 $\pm$ 16.54 \\
    \\
    BCG1& [-17.94] & [-19.96] & [0.37] & [14.80] & [0.15] & 44.38 $\pm$ 17.28 & 314.08 $\pm$ 11.45  \\
    \\
    BCG2& [0.00] & [0.00] & [0.20] & [-76.0] & [0.15] & 30.12 $\pm$ 8.42 & 413.92 $\pm$ 16.30  \\
    \\
    L\textsuperscript{*} Galaxy & - & - & - & - & [0.15] & 19.4 $\pm$ 5.2 & 191.5 $\pm$ 8.5 \\
    \hline\hline
    \end{tabular}}
\normalsize
\label{tab:a2744_sl}
\end{table*}


\label{lastpage}
\end{document}